\documentclass[sn-mathphys,Numbered]{sn-jnl}

\usepackage{enumitem}
\usepackage{graphicx}%
\usepackage{multirow}%
\usepackage{amsmath,amssymb,amsfonts}%
\usepackage{amsthm}%
\usepackage{amsfonts}
\usepackage{mathrsfs}%
\usepackage[title]{appendix}%
\usepackage{natbib}
\usepackage{xcolor}%
\usepackage{textcomp}%
\usepackage{manyfoot}%
\usepackage{booktabs}%
\usepackage{algorithm}%
\usepackage{algorithmicx}%
\usepackage{algpseudocode}%
\usepackage{listings}%
\newcommand{\overbar}[1]{\mkern 1.5mu\overline{\mkern-1.5mu#1\mkern-1.5mu}\mkern 1.5mu}

\newtheorem{theorem}{Theorem}
\newtheorem{proposition}[theorem]{Proposition}%

\begin{document}
	
	\title[Article Title]{Mathematical model of CAR-T-cell therapy for a B-cell Lymphoma lymph node}
	
	\author*[1]{\fnm{Soukaina} \sur{Sabir}}\email{Soukaina.Sabir@uclm.es}
	
	\author[1,2]{\fnm{Odelaisy} \sur{León-Triana}}
	
	\author[3]{\fnm{Sergio} \sur{Serrano}}
	
	\author[3]{\fnm{Roberto} \sur{Barrio}}
	
	\author[1]{\fnm{Victor M.} \sur{Pérez-García}}
	

	\affil*[1]{\orgdiv{Department of Mathematics}, \orgname{Mathematical Oncology Laboratory (MOLAB), Universidad de Castilla-La Mancha}, \orgaddress{\city{Ciudad Real}, \country{Spain}}}
	
	\affil[2]{\orgdiv{Translational Research in Pediatric Oncology, Hematopoietic Transplantation and Cell Therapy}, \orgname{Hospital La Paz Institute for Health Research-IdiPAZ}, \orgaddress{ \state{Madrid}, \country{Spain}}}
	
	\affil[3]{\orgdiv{Department of Applied Mathematics}, \orgname{Computational Dynamics Group (CoDy), Universidad de Zaragoza}, \orgaddress{ \city{Zaragoza},  \country{Spain}}}
	
	\abstract{CAR-T cell therapies have demonstrated significant success in treating B-cell leukemia in children and young adults. However, their effectiveness in treating B-cell lymphomas has been limited. Unlike leukemia, lymphoma often manifests as solid masses of cancer cells in lymph nodes, glands, or organs, making these tumors harder to access thus hindering treatment response. In this paper we present a mathematical model  that elucidates the dynamics of diffuse large B-cell lymphoma and CAR-T cells  in a lymph node. The mathematical model aids in understanding the complex interplay between the cell populations involved and proposes ways to identify potential underlying dynamical causes of treatment failure. We also study the phenomenon of immunosuppression induced by tumor cells and theoretically demonstrate its impact on cell dynamics. Through the examination of various response scenarios, we underscore the significance of product characteristics in treatment outcomes.}


	\maketitle
	
	\section{Introduction}\label{sec1}
	
	Non-Hodgkin's lymphoma (NHL) encompasses a diverse group of malignancies characterized by abnormal clonal proliferation of faulty immune system cells, either T-cells, B-cells, or both accounting for 3\% of cancer diagnoses globally \citep{Tandra}. In adults, the majority of NHL cases are of type B \citep{Sehn}. The most prevalent subtype among NHLs is the diffuse large B-cell lymphoma (DLBCL), with an incidence as high as 7.2 per 100,000 individuals per year \citep{Wang}.
	
	Immunotherapies have emerged as a highly effective treatment option for hematological malignancies, with Chimeric Antigen Receptor (CAR) T cell therapy being the most successful in use today. In these innovative therapies, first approved by the FDA only in 2017, T-cells are extracted from the patient's blood and undergo genetic engineering within a laboratory to introduce a chimeric antigen receptor (CAR) tailored to cancer cells. Subsequently, these modified T-cells are cultured and expanded, generating a robust population. Once infused back into the patient, the CAR-T cells recognize and bind to cancer cells by targeting specific proteins on their surface. This binding activates the CAR-T cells, initiating a robust immune response characterized by the release of cytotoxic substances, ultimately leading to the destruction of cancer cells \citep{Stein,Sterner}. This personalized immunotherapy has demonstrated remarkable success for adults with DLBCL \citep{Sheikh} as well as other haematological malignancies \citep{Zhang}, and even non-cancerous diseases \citep{Baker}. Promising results have been observed with high response rates and even complete responses sustained for long periods of time in DLBCL  \citep{Cappell}.
	CAR-T cells are currently offering a glimmer of hope for treating refractory or relapsed haematological cancers. However, several uncertainties remain, including identifying which patients will respond to treatment, how to sustain the treatment's effectiveness, and the long-term tolerance of CAR-T cells \citep{Zhang}. 
		
	Unlike other treatments with simple pharmacokinetics, CAR-T cells have a complex dynamics once infused. They first expand upon encountering their target antigen, in the case of haematological cancers most often CD19. This antigen is expressed both in cancer and healthy B-cells. After a fast initial expansion phase, the removal of the target cells leads to a decrease in the CAR-T population, also due to the exhaustion of T-cells. Thus, the use of mathematical models can be of value in comprehending the intricacies of CAR-T cell treatment and its effects. Indeed, many studies have employed diverse mathematical models to investigate various aspects of CAR-T cell therapies on different cancers \cite{ sahoo, victor, barros, juan, Russel, Russell2, depilis, barros, ivana, Liu, Ode, salvi, brain, alt}.
	
	Interestingly, despite the high incidence of lymphoma being the cancer type with more patients treated by CAR-T cells by far, the number of mathematical models addressing its treatments has been very small. Before the CAR-T era, \cite{roe} adapted Kuznetsov's model, originally developed to describe leukemia in mice, to describe the interaction between a proliferating Non-Hodgkin´s Lymphoma, an anti-tumor immune response, and the impact of chemotherapy on both the tumor and immune system. Notably, certain research findings indicate that more intensive therapies may lead to suboptimal tumor control. The objective of their work was to provide a plausible explanation for some of the paradoxical effects observed after chemotherapy therapy in lymphomas, while considering the significant anti-tumor role played by the immune system.
	
	In a recent study, \cite{alt} studied the dynamics and relationships between normal T-cells, CAR-T cells, and tumor cells in diffuse large B-cell lymphoma (DLBCL). The significance of this model lies in its depiction of cure as a stochastic event, highlighting the unpredictable nature of tumor eradication. This model can be applied to evaluate the effectiveness of CAR-T cell therapy in DLBCL and gain further insight into the impact of deterministic and stochastic factors on the outcomes. However, that model did not incorporate two relevant biological elements: the immunosuppression of T-cell action by both tumor cells and their microenvironment and the stimulation of CAR-T cell proliferation by the tumor antigens. These elements can profoundly impact in the treatment response and population dynamics in cases of diffuse large B-Cell lymphoma. Immunosuppression shapes the immune response and influences treatment outcomes, as highlighted by previous research \citep{brain, kuz}. Incorporating immunosuppression into the model is essential for a more realistic representation of the intricate interplay between the immune system and the lymphoma microenvironment, specially taking into account the compact nature of these tumors. Additionally, CAR-T cell proliferation after encountering tumor cells is a significant contributor to the efficacy of CAR-T cell therapy, as indicated by previous studies \citep{Ode, salvi,barros}, and it is certainly the leading contribution to expansion during the initial stages of response to treatment. Integrating all of these aspects into the modelling approach would enhance our ability to simulate and predict CAR-T cell behavior, providing a more comprehensive mechanistic understanding of the lymphoma response to these therapies.
	
	In this study, our goal is to address current knowledge gaps by crafting a simple, yet comprehensive, mathematical model. This model aims to clarify the dynamic interplay between CAR-T cells and Lymphoma B-cells, specifically within the intricate microenvironment of lymph nodes—also termed lymph node area. Lymph nodes, small bean-shaped structures integral to the lymphatic system, constitute a network of vessels and organs crucial for immune function. Distributed throughout the body and concentrated in specific regions like the neck, armpits, groin, chest, and abdomen, these regions collectively constitute what is commonly referred to as lymph node areas. The focus on this anatomical context enhances our ability to capture the nuanced complexities of the cellular dynamics characterizing the interaction between CAR-T cells and lymphoma B-cells within a lymph node area. We provide a comprehensive explanation of the assumptions, requirements, and equations utilized in our model. Furthermore, we derive the mathematical properties of the model and thoroughly discuss the plausibility of our proposed parameters.
	
	\section{Mathematical model}
	\label{Sec:MathModel}
	\subsection{Model Development}
	In this study, we build a mechanistic model able to describe mechanistically the response to CAR-T treatments of non-Hodgkin's B-cell lymphoma in a lymph node area. Our mathematical model accounts for the time-evolution of the basic interacting cellular populations grouped in two cellular compartments. Let us denote
 the number of CAR-T cells and lymphoma B-cells as $C(t)$ and $L(t)$ respectively, where $t$ stands for time. The system of differential equations governing the dynamics of these populations will be taken to be:
		\begin{subequations}\label{mainEqs}
		\begin{eqnarray}
			\frac{d C}{dt} & = & \sigma(t) +\beta\dfrac{CL}{H+L} - \gamma\dfrac{CL}{G+C} -\dfrac{C}{\tau_{C}}, \label{11}\\	
			\frac{d L}{dt} & = & \rho L- \alpha LC. \label{12}
		\end{eqnarray}
	\end{subequations}
The first term in Eq. \eqref{11} accounts for a external contribution to the CAR-T compartment given by the (in general) time dependent function $\sigma(t)$, that in our study will be assumed to be constant.  This term accounts for the fraction of CAR-T cells injected in the patient that get to the area under study but also on contributions from the expansion of the product in other body areas. This rate would depend also on the number of B-lymphocytes present in the patient's body after treatment. For example, if the patient undergoes lymphodepletion before CAR-T cell treatment, the external influx $\sigma$ of CAR-T cells getting to the lymph node area will be low due to the absence of B-cells stimulating CAR-T cell growth. In general, during the expansion phase, the contribution of this term will be negligible since most of the dynamics will be driven by local activation due to CAR-T cell interaction with the tumor antigen. However, when the antigen is depleted locally, the contribution of normal B-cell generation in the bone marrow will lead to some activation of the remaining CAR-T cells and will lead to a small constant flux through the body, that is accounted for in this term \citep{Serrano}. 
	
The second term in Eq. \eqref{11} accounts for the stimulation of CAR-T cell proliferation after their encounters with lymphoma cells. This term describes the rate at which CAR-T cells expand in the region of B-cell lymphoma due to the presence of tumor cells and is expected to be the main local contribution to the treatment expansion. It is important to point out that the rate of stimulation reaches a maximum value  when the number of lymphoma cells is large. The parameter $\beta$ measures the maximum mitotic stimulation after encounters with lymphoma cells, and $H$ measures the lymphoma cell population that provides half of the maximum stimulation, i.e. when $L=H$, stimulation is $\beta/2$. This type of terms have been used previously \citep{brain}, and are preferable to other similar stimulation terms without any saturation used in the literature \citep{rock,Ode,victor}. Unlike terms of the form $CL$, they provide a limitation on the maximum expansion rate that the product can have, what reflects the maximum rate at which CAR-T cells can perform mitosis. In this study we will not include an independent term accounting for the proliferation stimulation related to the release of cytokines during CAR-T cell stimulation. Those cytokines have an accumulative effect due to their finite residence time in the lymph node area and could be described in different ways, either through a nonlocal activation term in Eq. \eqref{11} or as an additional equation accounting the cytokine levels. Our choice here implicitly assumes that the residence time of cytokines is short so that the additional independent stimulation effect can be incorporated in the coefficient together with the direct stimulation $\beta$.

In tumors featuring a solid component, as it happens in lymph nodes, another key phenomenon is the induction of immune suppression by the tumor. This is delineated by the third term in Eq. \eqref{11}, signifying the deactivation of CAR-T cells by cancer cells. The maximum deactivation rate per cancer cell is denoted as $\gamma$, with the typical level of cellular saturation hovering around $G$. This type of terms have been employed previously in the context of CAR-T cell therapy against brain tumors \citep{brain}. Here we assume that all of the tumor cells exert an immunosupressive effect on CAR-T cells, in line with the assumption of well-mixed populations implicit in the compartimental approach. In compact solid tumors one could consider alternatively terms with powers of the tumour cell number, e.g. $T^{2/3}$, to account for the fact that only the most accessible tumor population will be able to suppress CAR-T cell activity. However, probably only spatial models can account for the complexity of the spatio-temporal dynamics of the whole tumor-immune cell interaction.
	
	The last term in Eq. \eqref{11} describes the natural death (or inactivation) of activated CAR-T cells, with a characteristic time $\tau_c$, depending typically on the CAR-T product properties and the persistence time of the activation of T-cells.
	
	Equation (\ref{12}) describes the dynamics of lymphoma cells. For simplicity we assume in the first term that the growth of the tumor cell population follows an exponential pattern with a growth rate $\rho>0$. While some studies have applied a logistic growth model to lymphoma cells \citep{gan,kuz}, we opted for a simpler approach by introducing an exponential growth term for the tumor \citep{roe,alt}. This decision is informed by the observation that, in the early stages of tumor growth, both models demonstrate comparable dynamics for lymphoma cells and at the stage at which the disease is detected and treated the tumor has not typically hit any anatomical barriers. Untreated real malignant cancers in humans have probably a faster growth \citep{perez}, but after treatment the remnant tumor clonal composition is substantially reduced so that lower powers are expected to rule tumor growth \citep{ocana}.  The exponential growth term provides both a simple description of growth with a minimal number of parameters and a balance between evolutionary forces and geometrical constraints that are present in the natural history of cancers and more specifically in lymphomas at treatment stage.

	The second term in Eq. (\ref{12}) accounts for the fact that CAR T-cells exert their anti-cancer effects through the law of mass action, where the killing rate is directly proportional to the product of the concentrations of CAR-T-cells and cancer cells \citep{brain,Ode, sahoo, Russel}.  The parameter $\alpha$ is related to the probability of an encounter between CAR-T and CD19$^{+}$ cells per unit of time and cell leading to the elimination of the target cell. This term does not include a saturation factor since, unlike mitotic processes that require substantially longer times to complete, encounters of T-cells with tumor cells can lead to a fast release of the cytotoxic load and elimination of the target in minutes \citep{dav}. Here we will consider that most of the death is due to those single-hit events, although recent evidences suggest that other more complex scenarios are possible \citep{Weigelin1,Weigelin2}. Various investigations have employed similar terms to characterize anti-tumor effects of T-cells \citep{Ode,salvi,brain,sahoo,roe,Russel}.
		
The biological factors governing the dynamics of CAR-T cells in our mathematical model include expansion and antigen stimulation, natural cell death, and inactivation by tumor cells.  The model should also integrate insights from CAR-T-cell therapy studies, emphasizing the crucial role of achieving sufficient lymphodepletion for a durable and effective treatment response. Consequently, patients undergo lymphodepleting chemotherapy either before or during CAR-T cell infusion, often resulting in a lack of normal B-cells. In line with \cite{alt}, we deliberately excluded the normal B-cell population from our model for simplicity, as their influence on the dynamics of CAR-T cells and lymphoma B-cells in the lymph node microenvironment is deemed negligible in the short and medium term due to the limited size of that population in patients.
	
In addition, our model aims to reflect that CAR-T cells can become inactivated  within the immunosuppressive tumor microenvironment. Indeed, large B-cell lymphoma originates from B lymphocytes, leading to the formation of solid masses or enlargements within the lymph nodes. The significant challenge in treating these cancers with CAR-T cells stems from the immunosuppressive tumor microenvironment (TME). This immunosuppression is orchestrated by various factors, including pro-tumor cell populations, cytokine profiles, metabolic immunosuppression, and vasculature, among others \citep{cheev}.	
	The obstacle presented by the immunosuppressive tumor microenvironment (TME) is a critical factor contributing to the failure of CAR-T-cell therapies for these tumors. Consequently, we have prioritized immunosuppression as a focal point in our mathematical model, aiming to attain a comprehensive understanding of cell dynamics in the context of CAR-T cell therapy for large B-cell lymphomas \citep{Sterner}.

	Previous studies on B-cell lymphoma have integrated the inactivation of immune cells upon exposure to tumor cells using mass-action terms \citep{kuz, ivana, roe, sant}. In our model, we incorporate a saturation factor to account for the saturation phenomenon, acknowledging that the effectiveness of immune cells may reach a plateau as they interact with tumor cells \citep{brain}. The incorporation of these biological effects within the mathematical model enhances its ability to capture the nuanced dynamics of immune responses in B-cell lymphoma patients.

	\subsection{Parameter estimation} \label{Parameter}
	
	The CAR-T inflow parameter $\sigma$, depends on the number of CAR-T cells present in the patient's body after treatment. In cases where patients undergo lymphodepletion before CAR-T cell therapy, the influx $\sigma$ into the tumor site situated within the lymph node tends to be diminished due to the absence of B-cells, which play a role in stimulating the proliferation of CAR-T cells. An estimation of effector cell production for human diffuse large B-cell lymphoma from the work of \cite{roe} is approximately $2 \times 10^5$ cells per day. Assuming that the biological mechanisms governing effector cells are similar to those of CAR-T cells, we consider a plausible range for $\sigma$ to be between $10^5$ and $10^7$ cells per day.
	
	Next, the maximum mitotic rate denoted as $\beta$, which is linked to the stimulation effect of T-cells through interaction with the target, depends on the characteristics of the CAR-T product. We chose the range for this parameter to be between 0.1 and 0.9 day$^{-1}$, in accordance with values reported in previous works \citep{Ode,Stein}, and aligning with the observation that stimulated CAR-T cells can undergo several mitotic divisions per day.
	
	For current CAR-T products, the mean lifetime $\tau_C$ of activated CAR-T cells typically falls within the range of 1 to 4 weeks. Lymphoma B-cells, being fast-growing malignant cancers, exhibit a proliferation rate $\rho$ on the order of several weeks, albeit with considerable individual patient variation \citep{fr, lang, tub}. Hence, we set $\rho$ within the range of 0.01 to 0.2 day$^{-1}$.
	
An approximate range for the parameter $H$ is between 10$^8$ and 10$^{10}$ cells. This range indicates the lymphoma cell population at which significant stimulation from CAR-T cells is observed. In the context of effector cells and tumor inactivation, the rate was estimated in mice and human diffuse large B-cell lymphoma and equals $3.422 \times 10^{-10}$ day$^{-1}$cell$^{-1}$, as indicated by \cite{roe} and \cite{kuz}. For CAR-T cells, we choose the maximum rate of tumor inactivation $\gamma$ within the range from 0 to 1 day$^{-1}$. The CAR-T concentration for half-maximal tumor inactivation, represented by $G$, takes values in the range of 10$^6$ to 10$^9$ cells.
	
	Furthermore, CAR-T cells exhibit killing efficiency against tumor cells, with values ranging from 10$^{-11}$ to 10$^{-9}$ day$^{-1}$cell$^{-1}$, as documented in \cite{Ode}.
	
	A comprehensive summary of the model parameters and their respective numerical values is presented in Table \ref{table1}.
	\begin{table}[h]
	\caption{Parameter values for Eqs.  \eqref{mainEqs} used in this work: names, description, values, units and sources.}
	\label{table1}
		\begin{tabular}{|c|l|c|c|c|}
			\hline& & & & \\
\textbf{Parameter}&\textbf{Description}&\textbf{Value}&\textbf{Units}& \textbf{Source}\\
			& & & & \\   \hline 			& & & & \\
			$\sigma$& External inflow of   & $10^{5}$--$10^{7}$	& 	cells$\times$ 	 &\cite{roe}\\ 			& CAR-T cells & & day$^{{-1}}$ & \\ 		&  & & & \\
\hline 			& & & & \\
			$\beta$& Mitotic stimulation  of & 	& 		& 		\cite{Stein}	     \\
			& CAR-T cell proliferation      &      0.1--0.9                                      &      day$^{{-1}}$                    &\cite{Ode}\\
			&   by tumor cells        &                                      &                                   &                       \\  
				&           &                                      &                                   &                       \\

					\hline 			& & & & \\

			$H$& Saturation to CAR-T & & 	& 		     \\
			&  cell stimulation rate   &             10$^{\text{8}}$--10$^{\text{10}} $                               &      	cells                      &                         \\
			&  &                                      &                                   &                       \\ \hline 			& & & & \\
			$\gamma$ &Tumor inactivation rate &      0--1 	& 	day$^{{-1}}$			 & 	\cite{kuz}\\ 			 & & & & \\  \hline 			& & & & \\

			$G$&  CAR -T inactivation   & & 		& 		     \\
			&    rate saturation constant  &                $ 10^{\text{6}}- 10^{\text{9}} $                            &             cells               &                   \\
					& & & & \\
 \hline 			& & & & \\

			$\tau_C$  & Activated CAR-T           & 7--14 			& days \hbox{ }\hbox{ } &\cite{Ode}\\
			& cell lifetime				     &						&				&			    \\ 			 & & & & \\
\hline 			& & & & \\

			$\rho$  & Tumor growth rate             &  0.01--0.2                		        & day$^{{-1}}$  &\cite{Ode}\\  			& & & & \\
\hline 			& & & & \\

			$\alpha$ & Killing efficiency  of &       	& 	  	day$^{{-1}}\times$		&  			     \\
			& CAR-T cells 			     &  10$^{-\text{11}}$-- 10$^{-\text{9}}$                                         &        cells$^{{-1}}$              &\cite{Ode}\\		& & & & \\
\hline
		\end{tabular}

	\end{table}

	\section{Equilibrium points and stability}
	\subsection{Non-dimensionalized model}	
	To study the equilibria and their stability in this section we will use a nondimensional form of Eqs. \eqref{mainEqs}, redefining the cell populations and time as follows
	
	\begin{equation}
		\overbar{C}= \frac{\rho}{\sigma} C, \; \;   \; \;\overbar{L}= \frac{L}{H},\; \;   \; \; \overbar{t}= \rho t.\label{relation}
	\end{equation}
	
	The non-dimensional parameters are related to the dimensional ones in the following way
	\begin{equation}
		m_1= \frac{\beta}{\rho} , \; \;    m_2= \frac{\gamma H}{\rho G},\; \;    m_3= \frac{\sigma}{\rho G}, \; \;
		m_4 = \frac{1}{\rho \tau_{C}}, \; \;    m_5 = \frac{\alpha \sigma}{\rho^2}.
	\end{equation}
	
	Then, our non-dimensionalized system becomes
	
	\begin{subequations} \label{NonDimen_model}
		\begin{eqnarray}
			\frac{d\overbar{C}}{d\overbar{t}} & = & 1 +m_1 \dfrac{\overbar{C}\overbar{L}}{1+\overbar{L}} - m_2\dfrac{\overbar{C}\overbar{L}}{1+m_3\overbar{C}} -m_4\overbar{C}, \label{NonDimen_model_1} \\	
			\frac{d \overbar{L}}{d\overbar{t}} & = & \overbar{L}- m_5\overbar{C}\overbar{L}. \label{NonDimen_model_2}
		\end{eqnarray}
	\end{subequations}
	To ensure the biological significance of this model, it is essential for the trajectories of the dynamical system described by Eq. \eqref{NonDimen_model} to be positively invariant. However, that follows easily from the facts that the axis $\overbar{L} = 0$ is invariant and that within the first quadrant, it is evident that $d\overbar{C}(t)/dt > 0$ when $\overbar{C}(t) \ll 1$. Thus, the following proposition holds:
	\medskip
	
		\begin{proposition}
		For any non-negative initial data given by $(\overbar{C_{0}},\overbar{L_{0}}) $, the trajectories of Eqs. \eqref{NonDimen_model} are positively invariant.
	\end{proposition}
	\medskip

	The existence and uniqueness of trajectories are direct consequences of the $\mathcal{C}^\infty$ nature of the second term of the differential system within the first quadrant.

	\subsection{Steady States and Stability Analysis}
	
	Although Eqs. (\ref{NonDimen_model}) are a pair of autonomous ODEs determining a planar dynamical system with a simple form,  its phase space has a complex structure. To understand the different dynamics that are possible in this system we will first study the nullclines and equilibrium points focusing in the regions of biological signifcance. Equilibrium points of biological significance are those where both $\overbar{C}$ and $\overbar{L}$ are non-negative. It is important to note that all of the model parameters must be non-negative to represent the intended biological phenomena.
	
	Equilibrium points for Eqs. (\ref{NonDimen_model}) are found at the intersections of nullclines, where the curves along which $\dot{\overbar{C}}=0$ and $\dot{\overbar{L}}=0$ intersect. The first equilibrium point, denoted as
	\begin{subequations} \label{equilibriaP}
	\begin{equation}
	E_{1}=\left(\frac{1}{m_4}, 0\right),
	\end{equation}
	 is determined by examining the intersection of $\dot{\overbar{C}}=0$ and $\overbar{L}=0$. This point consistently exists and is characterized as positive. Its stability, however, hinges on the specific values of the system parameters.
	\bigskip
	
	\begin{proposition} \label{stability_E_1}
		The tumor-free equilibrium point  $E_{1}$ is asymptotically stable if $m_5>m_4$, and unstable if $m_5<m_4$.
	\end{proposition}
	
	\begin{proof}
		The eigenvalues of Jacobian at the equilibrium point are $\lambda_{1} = -m_{4} $ and $\lambda_{2} = 1-m_{5}/m_{4}$. Thus, $E_{1}$ is asymptotically stable if $m_{4}<m_{5}$.
	\end{proof}
	
	Proposition \ref{stability_E_1} implies that it is possible to change the state of the system from the tumor bearing state to the tumor-free point. We will later make use of this result.
	
	In addition to $E_1$, for certain parameter sets, there are two additional equilibria,
	\begin{eqnarray}
	E_{2} & = & \left(\frac{1}{m_5}, \overbar{L}_2\right), \\
	E_{3} & = & \left(\frac{1}{m_5}, \overbar{L}_3\right),
	\end{eqnarray}
	\end{subequations}
	 corresponding to coexistence of the drug (CAR-T cells) with the tumor. In Eqs. (\ref{equilibriaP}b) and (\ref{equilibriaP}c),
	$\overbar{L}_2$ and $\overbar{L}_3$ are the solutions of
	\begin{equation}\label{roots_eq}
		a\overbar{L}^{2} + b\overbar{L}+c=0,
	\end{equation}
	with
	$$a=-\frac{m_2}{1+m_3/m_5}, \; \;  b= m_1+m_5-m_4- \frac{m_2}{1+m_3/m_5},\; \;  c=m_5-m_4.$$
	Therefore,
	\begin{subequations}
	\begin{eqnarray}
	 \overbar{L}_{2} & =&\dfrac{-b-\sqrt{\Delta}}{2a}, \label{Lbar2}\\
	\overbar{L}_{3} &=& \dfrac{-b+\sqrt{\Delta}}{2a}, \label{Lbar3}
	\end{eqnarray}
	\end{subequations}
	where $\Delta= b^2-4ac$.

Obviously, the existence of these two equilibria only occurs when $\Delta\geq 0$.
Let us see when this is the case.
	\bigskip
		
	\begin{proposition}\label{Prop_E23}
Equilibria $E_2$ and $E_3$ exist iff 		

\begin{itemize}
				\item $m_5\geq m_4$ or,
				\item $m_5< m_4$ and
    \begin{itemize}
      \item[$\circ$] $m_2 \leq \left(1+m_3/m_5\right)\left(\sqrt{m_{1}}-\sqrt{m_{4}-m_{5}}\right)^{2}$ or,
      \item[$\circ$] $m_2 \geq \left(1+m_3/m_5\right)\left(\sqrt{m_{1}}+\sqrt{m_{4}-m_{5}}\right)^{2}$.
    \end{itemize}
			\end{itemize}
	\end{proposition}
	\begin{proof}
	The discriminant is given by
				\begin{equation}\Delta= \left( m_1+m_5-m_4- \dfrac{m_2}{1+m_3/m_5}\right) ^2 + 4 \dfrac{m_2}{1+m_3/m_5}(m_5-m_4).\end{equation}
		If $m_5\geq m_4$, both summands are positive and therefore the discriminant $\Delta$ is positive. \\ 
		Now we have to analyse the case when $m_5 < m_4$. Defining $k=\frac{m_2}{1+m_3/m_5}$ we get $\Delta=k^2-2k(m_1+m_4-m_5)+(m_1-m_4+m_5)^2$. So $\Delta$ is positive if and only if $k\leq k_1$ or $k\geq k_2$, where $k_1=(m_1+m_4-m_5)-2\sqrt{m_1(m_4-m_5)}$ and $k_2=(m_1+m_4-m_5)+2\sqrt{m_1(m_4-m_5)}$ are the roots of the above polynomial. Substituting $k$ for its value and solving for $m_2$, we obtain the conditions of the statement.
\end{proof}
	
	To ensure that $E_2$ and $E_3$ hold biological significance, it is pertinent to investigate their positivity. Since $m_5>0$, the positivity of $E_{2}$ and $E_{3}$ hinges on the values of $\overbar{L}_{2}$ and $\overbar{L}_{3}$, respectively. The following proposition outlines the conditions that must be met for that to happen.
	\bigskip
		
	\begin{proposition}\label{Prop_positive}
		Assuming that $E_2$ and $E_3$ exist (see previous proposition):
		\begin{enumerate} 
			\item $\overbar{L}_2\geq 0 \Leftrightarrow$ \begin{itemize}
				\item $m_5\geq m_4$ or,
				\item $m_2 \leq (1+m_3/m_5)(m_1+m_5-m_4)$.
			\end{itemize} 
			\item $\overbar{L}_3\geq 0 \Leftrightarrow$ 	$m_2 \leq (1+m_3/m_5)(m_1+m_5-m_4)$ and  $m_5\leq m_4$.
		\end{enumerate}	
	\end{proposition}
	
	\begin{proof}
From Eq. (\ref{Lbar2}), $\overbar{L}_{2} \geq 0$ when the numerator is negative given that $a \leq 0$. To analyze the sign of the numerator, $-b-\sqrt{\Delta}$, we distinguish two cases. The first one corresponds to the case when $b\geq 0$ then $-b-\sqrt{\Delta} \leq0$ hence  $\overbar{L}_{2}$ is positive. Since $b= m_1+m_5-m_4- \frac{m_2}{1+m_3/m_5}$, this case occurs if and only if
$$m_2 \leq (1+m_3/m_5)(m_1+m_5-m_4).$$
In the second case, corresponding to
$b< 0$,
			$$\overbar{L}_{2} \geq 0 \Leftrightarrow -b-\sqrt{\Delta}\leq 0 \Leftrightarrow -b\leq \sqrt{\Delta} \Leftrightarrow b^2\leq\Delta \Leftrightarrow -4ac\geq 0 \Leftrightarrow  c\geq 0  \Leftrightarrow m_5\geq m_4.$$

		Now, let us focus on proving the second part of the proposition. To do so let us consider $\overbar{L}_{3}$, given by Eq. (\ref{Lbar3}). $\overbar{L}_{3}\geq 0 \Leftrightarrow -b+\sqrt{\Delta}$ is negative. This condition cannot hold if $b\leq 0$. Therefore, $b$ must be positive (i.e., $m_2 \leq (1+m_3/m_5)(m_1+m_5-m_4)$). Furthermore, $b$ must be larger than $\sqrt{\Delta}$. This leads to the inequality:

		$$\Delta\leq b^{2} \Leftrightarrow -4ac\leq 0 \Leftrightarrow c\leq 0 \Leftrightarrow m_5\leq m_4.$$
		
In agreement with our statement.
	\end{proof}
	
	It is worth noting that if $m_{5} = m_{4}$, we will have either $\overbar{L}_2=0$ (iff $b\leq 0$) or $\overbar{L}_3=0$ (iff $b\geq 0$). In such a scenario, the equilibrium point coincides with $E_1$, where the tumor cell population is reduced to zero while the CAR-T cells persist.
	
	Once the conditions for the positivity of the equilibria $E_2$ and $E_3$ have been stablished it is necessary to study the stability, since that would give us an idea of how biologically feasible it is to take the system to that state. To do so, we will calculate the possible local bifurcations that can occur for equilibria, thus getting the boundaries of the regions with different behaviour around them.

\subsection{Local bifurcations}
\label{sec:bif}
 Through bifurcation analysis \citep{guc, wigg}, we will explore how changes in the system parameters lead to different outcomes, such as stable coexistence, eradication of lymphoma cells, or immune escape.
\medskip
\begin{theorem}	
	\label{proptrans} The equilibria of Eqs.  \eqref{NonDimen_model} undergo a transcritical bifurcation at $m_{4}=m_{5}$:
	\begin{itemize}	
		\item When $m_2 > m_1(1+m_3/m_5)$, $E_{1}$ and $E_2$ experience a transcritical bifurcation.
		\item When $m_2 < m_1(1+m_3/m_5)$, $E_{1}$ and $E_3$ experience a transcritical bifurcation.
	\end{itemize}	
\end{theorem}
	\begin{proof}
According to Proposition \ref{stability_E_1}, the stability of the equilibrium $E_1$ changes when $m_4=m_5$. On the other hand, $E_1$ and $E_2$ coincide when $m_4=m_5$ and $m_2 \geq m_1(1+m_3/m_5)$. Whereas $E_1$ and $E_3$ coincide when $m_4=m_5$ and $m_2 \leq m_1(1+m_3/m_5)$. Finally, since the Jacobian matrix for $E_i$, with $i=2,3$, is
	\begin{equation} \label{Jacob} J\left( \frac{1}{m_{5}}, \overbar{L}_{i}\right) =\begin{pmatrix}
		\dfrac{m_{1}\overbar{L}_{i}}{1+\overbar{L}_{i}}-\dfrac{m_{2}\overbar{L}_{i}}{(1+\frac{m_3}{m_5})^{2}}-m_{4}  &\phantom{m}&  \dfrac{m_{1}}{m_{5}(1+\overbar{L}_{i})^{2}}-\dfrac{m_{2}}{m_5(1+\frac{m_3}{m_5})}\\[3ex]
		-m_5 \overbar{L}_{i}&& 0
	\end{pmatrix},\end{equation}
it is clear that its determinant changes sign when $\overbar{L}_i$ changes sign (note that when $\overbar{L}_{i}=0$, $J_{1,2}$ changes sign only when $m_2=m_1(1+m_3/m_5)$). This happens in the cases described above.
\end{proof}

\noindent
{\bf Remark:} In both cases the equilibria of Eqs. \eqref{NonDimen_model} go from $E_1$ stable and $E_i$ saddle when $m_5>m_4$ to $E_1$ saddle and $E_i$ stable when $m_5<m_4$, where $i$ is 2 or 3, depending on the value of $m_2$.

As discussed above $E_2$ and $E_3$ exist only under certain parametric conditions (see Proposition \ref{Prop_E23}). The boundary between the different regions of existence for $E_2$ and $E_3$ is determined by a fold bifurcation.

\medskip
\begin{theorem}	
	\label{propfold} The equilibria $E_2$ and $E_3$ of the dynamical system \eqref{NonDimen_model} undergo a fold bifurcation when $m_5\leq m_4$ and $m_2 = \left(1+m_3/m_5\right)\left(\sqrt{m_{1}}\pm\sqrt{m_{4}-m_{5}}\right)^{2}$.
\end{theorem}
	\begin{proof}
As stated in the proof of Proposition \ref{Prop_E23}, the value of $\Delta$ cancels out when $m_5\leq m_4$ and $m_2=\left(1+m_3/m_5\right)\left(\sqrt{m_{1}}\pm\sqrt{m_{4}-m_{5}}\right)^{2}$. In this situation, $E_2=E_3$; in the outer region, $E_2$ and $E_3$ exist and are distinct; and in the inner region, $E_2$ and $E_3$ do not exist.
\end{proof}

\noindent
{\bf Remark:} If we fix the values of $m_1$, $m_3$ and $m_4$, the condition $m_2 = \left(1+m_3/m_5\right)\left(\sqrt{m_{1}}\pm\sqrt{m_{4}-m_{5}}\right)^{2}$ (with $m_5\leq m_4$) determines a curve with two branches. The branches $m_2 = \left(1+m_3/m_5\right)\left(\sqrt{m_{1}}-\sqrt{m_{4}-m_{5}}\right)^{2}$ and $m_2 = \left(1+m_3/m_5\right)\left(\sqrt{m_{1}}+\sqrt{m_{4}-m_{5}}\right)^{2}$ meet at the point $m_5=m_4$ and $m_2=m_1(1+m_3/m_5)$. At this point the derivative of $m_5$ with respect to $m_2$ is well defined and is 0 (i.e. the line $m_5=m_4$ is tangent to the curve at this point). Note also that at this point the three equilibria are equal, and if we pass through it along any straight line in the decreasing direction of $m_5$, the dynamical system goes from having three equilibrium points (with $E_1$ stable and the other two unstable) to a single equilibrium point, $E_1$, which becomes unstable, thus this point is a subcritical pitchfork bifurcation point.

\medskip
\begin{theorem}	
	\label{propfold2} The equilibria $E_1$, $E_2$ and $E_3$ of the dynamical system \eqref{NonDimen_model} undergo a subcritical pitchfork bifurcation when $m_5= m_4$ and $m_2=m_1(1+m_3/m_5)$.
\end{theorem}
	\begin{proof}
See previous remark.
\end{proof}

\noindent
{\bf Remark:} Once $m_1$, $m_3$ and $m_4$ are taken to be fixed and positive, $\forall\, 0<m_5\leq m_4$ it is clear that $m_2\geq 0$ along the fold bifurcation: If $m_5\rightarrow 0^+$, $m_2\rightarrow +\infty$ on both branches. On the other hand, the derivative of $m_2$ with respect to $m_5$ on the left branch cancels out only if

\medskip
\begin{itemize}
\item  $m_5=m_4- m_1$ (which implies $m_2=0$) when $m_1<m_4$,
 \item $m_1=\dfrac{(m_3m_4+m_5^2)^2}{m_3^2(m_4-m_5)}$ (which implies $m_2=\dfrac{(m_3+m_5)^2}{m_3^2(m_4-m_5)}>0$) when $m_1>m_4$.
      \end{itemize}
      \medskip
    
The pitchfork bifurcation is also present in the positive parametric region, since it lies within the fold bifurcation.

\begin{theorem}	
	\label{propHopf} A Hopf bifurcation occurs for $E_3$ when 
	\begin{equation}
	m_2=\dfrac{(m_3+m_5)^2(m_1m_3-m_4m_3-m^2_5)}{m_3(m_3m_4+m^2_5)},
	\end{equation}
under the following conditions:
\begin{itemize}
  \item $m_1>\dfrac{(m_3m_4+m_5^2)^2}{m_3^2(m_4-m_5)}$,
  \item $m_4>m_5$.
\end{itemize}
\end{theorem}
	\begin{proof}
		The Jacobian Eq. \eqref{Jacob} exhibits purely imaginary eigenvalues when $J_{1,1}=0$ and $J_{1,2}\cdot J_{2,1}<0$.		
		 Let us begin by defining $\overbar{L}_{i}$ as function of the parameters of the system (\ref{NonDimen_model}). To accomplish this, we examine the following equation given by $J_{1,1}=0$:
		 $$\dfrac{-m_2}{(1+m_3/m_5)^2}\overbar{L}^2_{i}+\overbar{L}_{i}\left( m_1-m_4-\dfrac{m_2}{(1+m_3/m_5)^2}\right) -m_4=0.$$
		 We multiply the equation above by
		 $(1+m_3/m_5)$ and then calculate its difference with \eqref{roots_eq}, yielding:
		 $$\overbar{L}_i=\overbar{L}_H=\dfrac{m_3 m_4 +m^2_5}{m_1 m_3 -m_3 m_4 -m^2_5}.$$
		 Substituting $\overbar{L}_H$ into  $J_{1,1}=0$, we obtain the first condition for the Hopf bifurcation:
		 $$m_2=\dfrac{(m_3+m_5)^2(m_1m_3-m_4m_3-m^2_5)}{m_3(m_3m_4+m^2_5)}.$$		
		
With these two conditions (for $\overbar{L}_H$ and $m_2$), substituting we find that $\overbar{L}_H$ corresponds to $\overbar{L}_3$. Now we need to ensure that the determinant of the Jacobian matrix is positive:
$$\overbar{L}_H\left(\dfrac{m_1}{(1+\overbar{L}_H)^2}-\dfrac{m_2}{1+m_3/m_5}\right)>0.$$
	 Substituting the explicit forms of $\overbar{L}_H$ and $m_2$, we get the Hopf bifurcation condition, 
	 
	 \begin{equation}
	 m_1>\dfrac{(m_3m_4+m_5^2)^2}{m_3^2(m_4-m_5)} \ , \ m_4>m_5.
	 \end{equation}
\end{proof}

\noindent
{\bf Remark:} The condition $m_1>\dfrac{(m_3m_4+m_5^2)^2}{m_3^2(m_4-m_5)}$ implies that $m_1>m_4$. So, if $m_1\leq m_4$, there is no Hopf bifurcation. In addition, the intersection between the fold bifurcation and the Hopf bifurcation occurs when $m_1=\dfrac{(m_3m_4+m_5^2)^2}{m_3^2(m_4-m_5)}$ that matches the backward (leftmost) point of the fold bifurcation (with $m_2=\dfrac{(m_3+m_5)^2}{m_3^2(m_4-m_5)}>0$). This codimension-two point (a Bogdanov-Takens bifurcation point) is the origin of the Hopf bifurcation curve. 

From the practical point of view, the existence of these bifurcations lead to qualitative changes in the solutions of Eqs.  \eqref{NonDimen_model}, thus having a substantial influence on the populations of lymphoma cells and CAR-T cells.

It is relevant to study the changes of $E_i$ from being of type focus to node, and vice versa. These are not bifurcations since there is no change in the stability, but those changes have implications in the form the solutions converge (or diverge) to (from) equilibria. The analysis is developed in Appendix \ref{AppA}. 

\subsection{The inflow of CAR-T cells into the lymph node area $\sigma(t)$ has a major role on the stability of the equilibria}

\begin{figure}[H]
	\centering
	\includegraphics[width=\columnwidth]{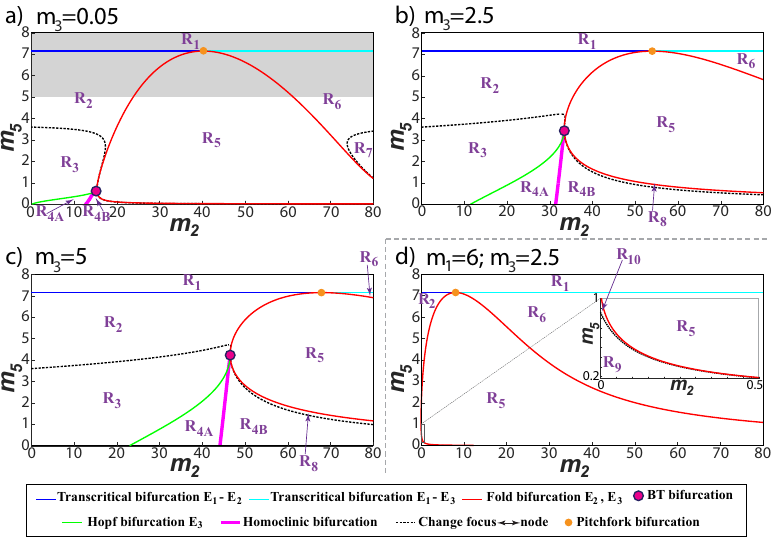}
	\caption{Different regions of behavior, labeled $R_1-R_9$, depending on the parameter values $m_2, m_5$. Three values for $m_3$ are shown (corresponding, for instance to different values of the influx of CAR-T cells into the tumor areas $\sigma$): a) $m_3= 0.05$, b) $m_3= 2.5$ and c) $m_3= 5$. The fixed values of the other dimensionless parameters are $m_1= 40$ and $m_4= 7.14$.   d) Example with $m_1=6, m_3 = 2.5$ corresponding to a case with $m_1=6<m_4$. (See Table \ref{tab_regions} for explanation about different regions $R_i$). The shaded area in subplot a) lies out of the range of biological interest identified in Table \ref{table1}. }
	\label{QualitativeRegions}
\end{figure}

Figure \ref{QualitativeRegions} shows the regions with qualitatively distinct dynamics as functions of the parameters $m_2$ and $m_5$ using the explicit formulas obtained in Sec. \ref{sec:bif}.  The figure shows also the homoclinic bifurcation curve obtained with the AUTO continuation software \cite{AUTO1,AUTO2}. As discussed in Sec. \ref{sec:bif}, the origin of this bifurcation curve is a Bogdanov-Takens (BT) bifurcation point, also shown in Fig. \ref{QualitativeRegions}  and determined analytically after Theorem \ref{propHopf}.

\begin{table}\centering
\begin{tabular}{|l|c|c|c|}\hline
&\multicolumn{3}{|c|}{Equilibria}\\\hline
Region&$E_1$&$E_2$&$E_3$\\\hline
$R_1$&{\bf biological significance}&biological significance&no biological significance\\
& {\bf attractor node}&saddle&saddle\\\hline
$R_2$&biological significance&biological significance&{\bf biological significance}\\
& saddle&saddle&{\bf attractor node}\\\hline
$R_3$&biological significance&biological significance&{\bf biological significance}\\
& saddle&saddle&{\bf attractor focus}\\\hline
$R_4$&biological significance&biological significance&biological significance\\
& saddle&saddle&repeller focus\\\hline
$R_5$&biological significance&$\nexists$&$\nexists$\\
& saddle&&\\\hline
$R_6$&biological significance&no biological significance&no biological significance\\
& saddle&attractor node&saddle\\\hline
$R_7$&biological significance&no biological significance&no biological significance\\
& saddle&attractor focus&saddle\\\hline
$R_8$&biological significance&biological significance&biological significance\\
& saddle&saddle&repeller node\\\hline
$R_9$&biological significance&no biological significance&no biological significance\\
& saddle&saddle&attractor focus\\\hline
$R_{10}$&biological significance&no biological significance&no biological significance\\
& saddle&saddle&attractor node\\\hline
\end{tabular}
\caption{Characteristics of the three equilibria of the system for the different regions located in Figure \ref{QualitativeRegions}.}\label{tab_regions}
\end{table}

Figure \ref{QualitativeRegions}a)-c) shows the regions of stability in the $m_2, m_5$ plane for three specific choices of $m_3$. The values chosen for $m_3$ can be obtained by changing the influxes $\sigma$ of CAR-T cells into the tumor localization. For instance Figure \ref{QualitativeRegions}a) corresponds to $\sigma=10^5$ (lower end of the interval given in Table \ref{table1}); Figure \ref{QualitativeRegions}b) to $\sigma=5\times 10^6$; and Figure \ref{QualitativeRegions}c) to $\sigma=10^7$ (upper end of the interval). In all cases we observe a similar structure of regions but as $m_3$ (i.e. $\sigma$) increases, the regions $R_2$, $R_3$ (and $R_4$) grow, shifting regions $R_5$-$R_8$ to the right. This has biological implications since $R_{2,3}$ correspond to the controlled tumor and CAR-T equilibria thus making clear that the maintenance of a flow of CAR-T cells in the system may have a positive effect on maintaining the disease under control. Also, it is clear that achieving a complete cure is only possible in this case for large values of $m_5$ that can be achieved also by increasing $\sigma$. Thus, this appears to be a key parameter able to drive the system in either tumor-free or tumor-controlled scenarios.

Figure  \ref{QualitativeRegions}d) shows the case with $m_3=2.5$ and $m_1=6$ in which there are new regions because $m_1<m_4$ (see Sec. \ref{sec:bif}). Now the region $R_5$ is tangent to the axis $m_2=0$ and regions $R_{3,4}$ do not exist. Instead the regions $R_9$ and $R_{10}$ appear. The description of the type of equilibria appearing in each region is listed in Table \ref{tab_regions}.

\begin{figure}
	\centering
	\includegraphics[width=0.92\columnwidth]{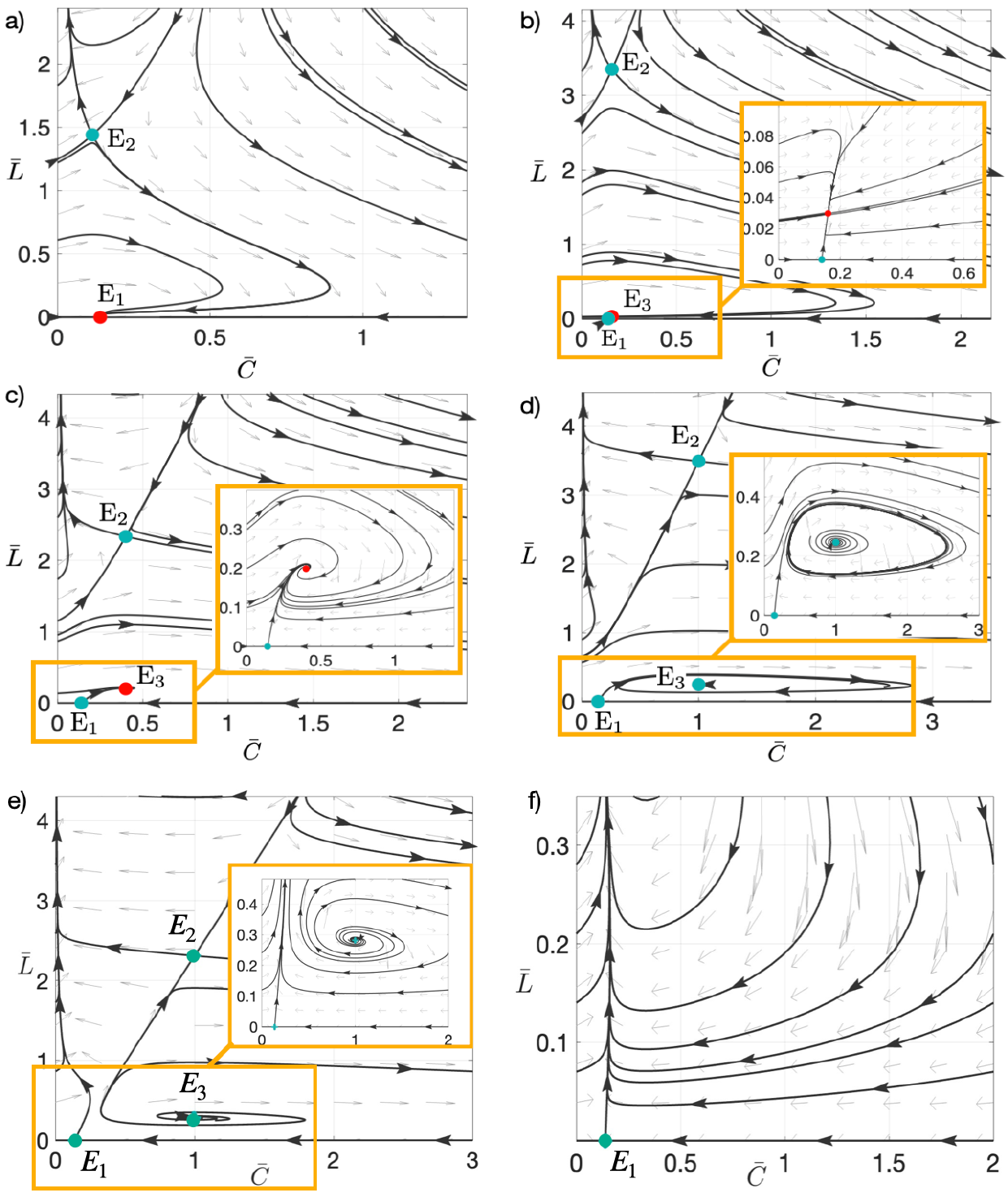}
	\caption{Phase portraits for the solutions of Eqs. \eqref{NonDimen_model} for different values of the parameters $(m_2, m_5)$. Subplots (a)-(f), correspond to pairs of values (22, 8.75), in $R_1$; (12.5, 6.25), in $R_2$; (20, 2.5), in $R_3$; (25, 1), in $R_{4A}$; (33, 1), in $R_{4B}$; and (40, 3), in $R_5$  respectively. With these choices, stability of the equilibria fall in the different regions $R_1-R_5$ discussed in the main text.  Other parameters where fixed to  $m_1= 40, m_3= 2.5$ and $m_4= 7.14$, corresponding with Fig. \ref{QualitativeRegions} b).}
	\label{SpaceFhase1}
\end{figure}	

\subsection{Parameters, initial tumor size and number of CAR-T cell injected determine tumor control}
\label{PhasSpa}
\begin{figure}[H]
	\centering
	\includegraphics[width=\textwidth]{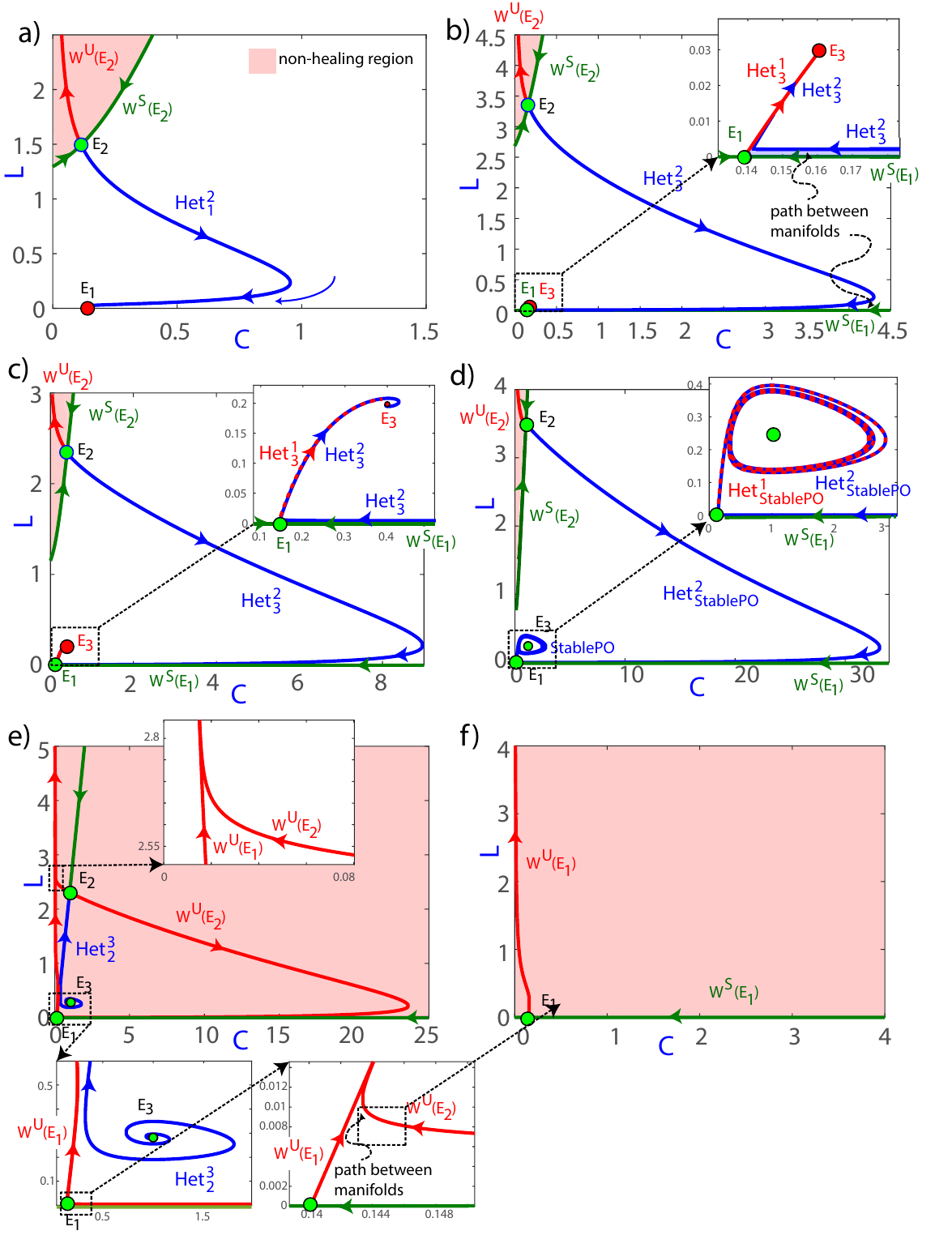}
	\caption{Stable ($W^s(E_i)$) and unstable ($W^u(E_i)$) manifolds of equilibria on the phase portraits for the solutions of Eqs. \eqref{NonDimen_model} for different values of the parameters $(m_2, m_5)$. Subplots (a)-(f), correspond to pairs of values (22, 8.75) in the $R_1$ region; (12.5, 6.25) in $R_2$; (20, 2.5) in $R_3$; (25, 1) in $R_{4A}$; (33, 1), in $R_{4B}$, and (40, 3) in $R_5$, respectively. Other parameters were fixed to  $m_1= 40, m_3= 2.5$ and $m_4= 7.14$. The heteroclinic (Het${}_i^j$) connections among equilibria or the stable periodic orbit are also shown. }
	\label{mandef}
\end{figure}

Representative phase portraits for different pairs of parameters $(m_2,m_5)$ corresponding to the different regions $R_1-R_5$ are shown in Figure \ref{SpaceFhase1}. The value of $m_3=2.5$ has been fixed as the value of Fig. \ref{QualitativeRegions} b) and other parameters listed in the caption. To study the phase space images in detail we have obtained the stable and unstable manifolds of the saddle equilibria in Fig.~\ref{mandef}. Both figures are complementary and each plot of both figures corresponds to the same case (same parameter values and therefore same region). Therefore, we will comment on both figures in parallel.

The three steady states, denoted as $E_1$, $E_2$, and $E_3$, are marked with coloured circles (red for stable equilibrium and green for unstable ones).  It can be seen that in Fig.  \ref{SpaceFhase1}a (region $R_1$) only $E_1$ and $E_2$ are positive and just $E_1$ is stable. The basin of attraction of the equilibrium $E_1$  occupies a substantial area in the phase space, but there are orbits that are not bounded, corresponding to loss of control of the tumor (see Fig. \ref{mandef} for the stable and unstable manifolds of the equilibria and the shaded regions in which the tumor escapes the CAR-T surveillance).  $E_2$ is a saddle and its stable manifold ($W^S(E_2)$, shown in green) forms the boundary of the unbounded region (the non-healing region). There is an heteroclinic connection from $E_2$ to $E_1$. This cycle also gives a  separation in the bounded region, above the heteroclinic cycle the orbits decrease in $\overbar{L}$ going to large values of $\overbar{C}$ and approaching $\overbar{L}=0$, and converging to $E_1$. In the remaining area the orbits converge faster to $E_1$. In practice unbounded orbits are all located in a region with large initial tumor loads and smaller numbers of CAR-T cells infused. This interesting result is fully in line with experimental observations where it is known that factors associated with durable remission after CAR-T cell therapy include lower baseline tumour volume and higher peak circulating CAR-T cell levels \citep{Cappell}. Similar observations apply in different scenarios  when $E_1$ is unstable but the coexistence equilibrium $E_3$ is stable as in Figs. \ref{SpaceFhase1}b,c (see explanation below).

Thus even in a situation where parameters could achieve a cure, the initial situation may have a substantial effect on the outcome, i.e. both parameter values and initial conditions have to be properly engineered to control the system. 

It is clear that when the parameter $m_5$ decreases, which can be associated with the killing efficiency rate $\alpha$, and $m_2$ is sufficiently low ($m_2<m_1(1+m_3/m_4)$), $E_1$ becomes unstable, and $E_3$ appears in the positive quadrant. We have crossed the transcritical bifurcation curve (see Theorem \ref{proptrans}). Remarkably,
	$E_3$ becomes a stable node, as illustrated in Fig. \ref{SpaceFhase1}b) (region $R_2$). The situation is quite similar, but a bit more complex, to the one depicted in Fig. \ref{SpaceFhase1}a). Now the heteroclinic cycles connect $E_2$ and $E_3$, and $E_1$ and $E_3$. This gives rise to a rather thin path between the stable manifold of $E_1$ ($W^S(E_1)$) and the heteroclinic cycle between $E_2$ and $E_3$. Once an orbit enters its interior, it passes through it towards $E_3$. The interesting fact is that, although the orbit goes to an endemic equilibrium, the orbit reaches very small values of $\overbar{L}$ for a long time before going to $E_3$. This explains several cases of disease relapse, but also allows us to design strategies to control it before it grows again. In Fig. \ref{SpaceFhase1}c) (region $R_3$), $E_3$ is depicted as a stable spiral, reflecting the evolution of its stability characteristics under varying conditions. This stable spiral signifies a more intricate dynamical behavior compared to the stable node observed previously. That is, the only difference is that the orbits oscillate before reaching $E_3$.

\begin{figure}[h]
	\centering
	\includegraphics[width=1\textwidth]{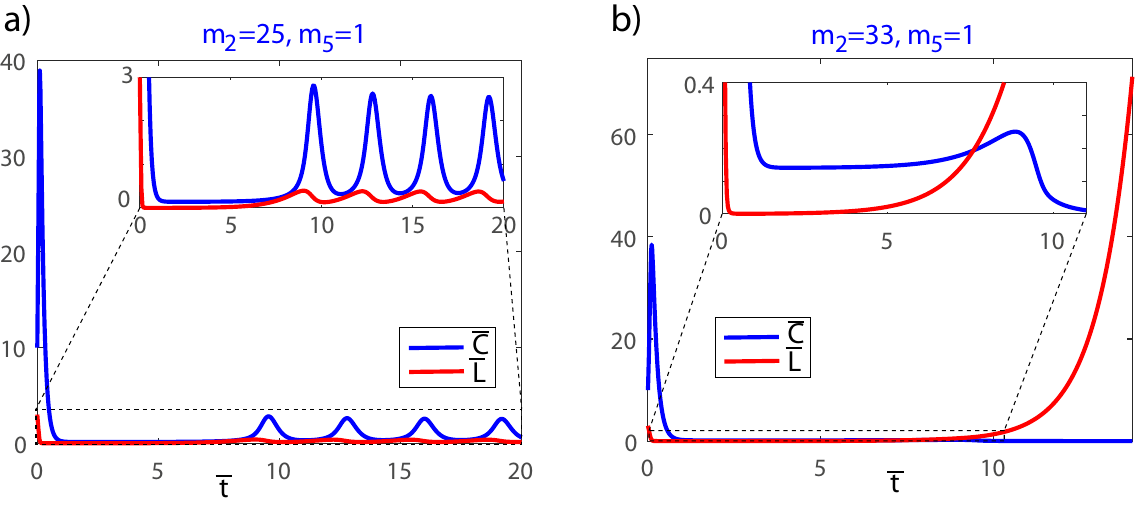}
	\caption{Examples of the time evolution of orbits in the $R_4$ region on both sides of the homoclinic bifurcation: a) $(m_2, m_5)=(25, 1)$ in $R_{4A}$ showing disease control  and b) (33, 1) in $R_{4B}$ displaying tumor escape. Both orbits have the initial conditions $(C(0), L(0))= (10, 3)$. }
	\label{ORBITASv2}
\end{figure}	

The behavior of $E_3$ undergoes a significant change when $m_5$ decreases and $m_2$ increases, as depicted in Fig. \ref{SpaceFhase1}d) (region $R_{4A}$). In this case the $E_3$ equilibrium point has undergone a supercritical Hopf bifurcation (see Theorem \ref{propHopf}). Under such conditions, $E_3$ transforms into a repeller spiral steady state, and so, the convergence is towards the stable periodic orbit around $E_3$ generated in the Hopf bifurcation. Note that the heteroclinic cycles lie between equilibria and the stable periodic orbit, but the rest of comments are the same as in plot c). In this region, as $m_2$ increases and we approach the homoclinic bifurcation curve, the period of the stable limit cycle increases until, upon reaching the homoclinic bifurcation, it becomes a homoclinic orbit of $E_1$. This is why the limit cycle has disappeared in the $R_{4B}$ region. This bifurcation is quite important and results in a global change to the system. This is the case shown in plot e). Now all the orbits are unbounded, that is, there is no cure or control (if no action is taken). The orbits on the right side can give small values of $L$ but then they enter into two paths between manifolds escaping through them giving rise to unbounded orbits. It is interesting to remark that the escape always follows a nearly vertical line at $C\thickapprox 0$.  In Figure~\ref{ORBITASv2} we present the time dynamics of two orbits in the $R_4$ region: plot a) with $(m_2, m_5)=(25, 1)$ in $R_{4A}$, and plot b) (33, 1) in $R_{4B}$. Both orbits have the initial conditions $(\overbar{C}(0), \overbar{L}(0))= (10, 3)$ which correspond to a point above the heteroclinic orbit and to the right of the stable manifold of $E_2$. In the pictures we can clearly see how after the homoclinic bifurcation (Fig. \ref{ORBITASv2}b) the orbit becomes unbounded, whereas in Fig. \ref{ORBITASv2}a) the orbit converges to the stable periodic orbit. Note that both orbits stay close to $E_1$ for some time (quite long, depending on the value of $\rho$, in case Fig. \ref{SpaceFhase1}b), meaning that the disease can be controled for long periods of time with small values of $\overbar{L}$.

Figs. \ref{SpaceFhase1}f) and \ref{mandef}f show typical scenarios in $R_5$ where there is a saddle equilibrium $E_1$, and all orbits escape. Orbits in regions $R_6$, $R_7$, $R_9$ and $R_{10}$ (not shown)  have a similar behavior to those in $R_5$, since both $E_2$ and $E_3$ equilibria exist, but they are not positive. Therefore, the only equilibrium in this quadrant is $E_1$. Since $E_1$ is of saddle type in these regions, the dynamics is not bounded and the cancer grows without control. The dynamics in $R_8$ is similar to that in $R_{4B}$, with the only difference that the equilibrium $E_3$ is a (repeller) node instead of a focus, thus all orbits are unbounded.

In summary, within $R_1$, $R_2$ and $R_3$ there is only a stable positive equilibrium point. The value of $\overbar{L}$ in those equilibria is either 0 (in $R_1$) or relatively small (in $R_2$ and $R_3$).  Thus, it may be possible to cure or take the disease to a stable state controlled by a remnant of CAR-T cells.
The equilibrium $E_2$ is unstable in those regions but determines the dynamics in a part of the phase space. Its stable manifold delimits the bounded region, and above this manifold, the orbits become unbounded, leading to uncontrolled tumor growth. To control the disease, trajectories must reside either within the basin of attraction of $E_1$ (in $R_1$) or $E_3$ (in $R_{2,3}$). It is important to note that equilibrium $E_1$ is predominantly unstable, except for $m_5>m_4$, which implies a scenario where CAR-T cells exert significant cytotoxic effects on tumor cells, surpassing the rate of tumor growth and potentially resulting in tumor regression or stabilization over time. When $E_1$ is unstable, the only control scenario implies converging towards the stable coexistence equilibrium $E_3$. Regions $R_{2,3}$ increase as $\sigma$ increases, so a larger inflow of CAR-T cells would make it easier to achieve tumor control. Within regions with controlled tumor growth, the eventual convergence of trajectories towards equilibria (either $E_1$ or $E_3$) depends on the specific initial conditions of the system, i.e. the initial number of CAR-T cells and the initial tumor load.
In the region $R_{4A}$, although $E_3$ is unstable, it is still possible to control the disease due to the existence of an attracting limit cycle generated in the Hopf that delimits this region. This implies that at the time of diagnosis, if the number of lymphoma cells and CAR-T cells are close to the $E_3$ steady state, it is possible to define a therapeutic protocol capable of taking the patient to a stable disease state. Thus, the model can explain both tumor dormancy and escape from CAR-T cells.

As the value of $m_2$ increases, indicating an escalation in the tumor inactivating rate $\gamma$, in the region $R_{4B}$, the stable periodic orbit abruptly disappears, consequently leading to an irrevocable non-cure situation. This observation underscores the critical importance and profound impact of the immunosuppressive tumor microenvironment in determining the success or failure of CAR-T cell treatment strategies. It also highlights the intricate interplay between the tumor microenvironment and the efficacy of CAR-T cell therapy, emphasizing the need for a comprehensive understanding of the complex dynamic interactions to enhance treatment outcomes.

\section{Study of the dynamics and implications for therapy}

In what follows we will work on the dimensional form Eq. (\ref{mainEqs})  of our model in order to  facilitate a more insightful analysis and interpretation of our findings.

\subsection{Correlation between killing efficiency and tumour inactivity rates with initial conditions}

\begin{figure}[H]
\centering	
\includegraphics[width=\columnwidth]{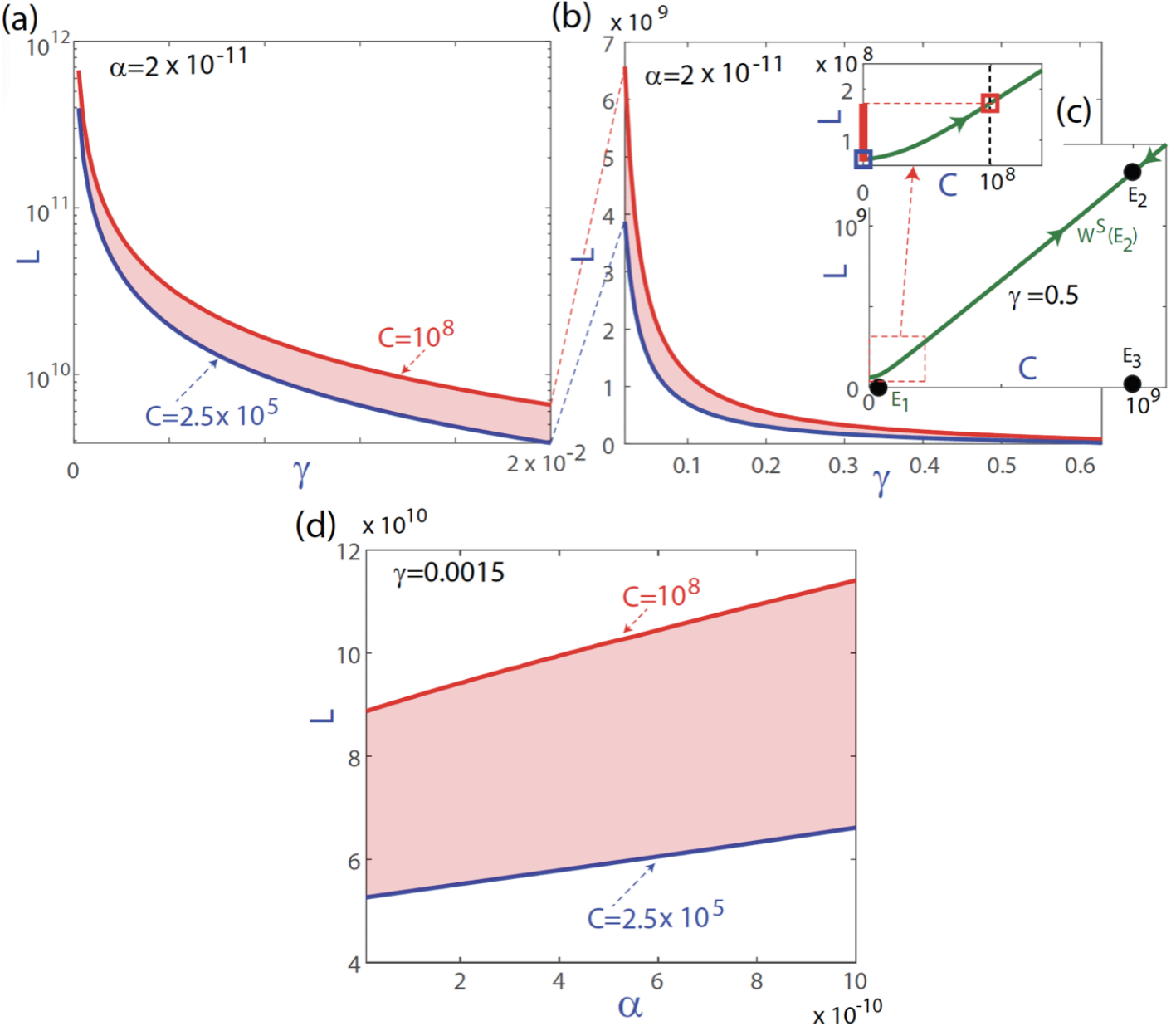}
\caption{Ranges of $L$ for which control can be achieved depending on $\gamma$ and $\alpha$ values. Parameters are $\beta=0.8$ day$^{-1}$, $H=G=10^{8}$ cells, $\tau_C=7$ days, $\sigma=5\times 10^{6}$ cells day$^{-1}$ and $\rho=0.02$ day$^{-1}$. (a,b) changing $\gamma$ and fixing $\alpha=2\times 10^{-11}$ day$^{-1}$. (c) Graphical explanation of the meaning of the blue and red curves. (d) Results varying $\alpha$ and fixing $\gamma= 0.0015$ cells$^{-1}\times$ day$^{-1}$.}
\label{limits}
\end{figure}

To explore the relation between the parameters  $\alpha$ and $\gamma$ and the initial conditions, in Figure \ref{limits} we show different curves obtained by varying the killing efficiency rate $\alpha$ or the tumour inactivity rate $\gamma$. The blue curves show the cutoff value of the stable manifold of the equilibrium $E_2$ (boundary between the basin of attraction of the stable invariant set and the region of unbounded dynamics) with the vertical line $C=2.5\times 10^5$. The red curves show the cutoff value of the manifold with the straight line $C=10^8$. Values of $L$ below the blue curve converge to the stable invariant for any value of $C\geq 2.5\times 10^5$. Values of $L$ above the red curve would require an initial injection of CAR-T above $10^8$ cells, that is a large value difficult to reach in realistic scenarios. In other words, the disease would have reached a point at which the therapy would no longer be effective. Between the two curves (coloured area), CAR-T therapy will or will not be effective depending on the initial dose injected. As it can be seen in Fig. \ref{limits}a), small values of $\gamma$ would allow the therapy to be effective even with high initial values of $L$. In contrast, as $\gamma$ grows (Fig. \ref{limits}b), larger values of $L$ can quickly prevent the therapy from being effective. The dependence on the $\alpha$ parameter (plot (d)) is essentially linear.

\subsection{Short-term and long-term effects of the initial tumor and CAR-T cell numbers on the dynamics}

The phase space analysis of Sec. \ref{PhasSpa} has allowed us to study the substantial influence of initial data $C(0), L(0)$ on the disease outcome. We move on now to study the time evolution for different sets of parameters and initial conditions. Let us first consider a tumor inactivation rate $\gamma=0.004$ day$^{-1}$ and a CAR-T kill rate $\alpha =2\times 10^{-10}$ day$^{-1}$ cells $^{-1}$. The remaining parameters were taken to be $\beta=0.8$ day$^{-1}$, $H=G=10^{8}$ cells, $\tau_C=7$ days, $\sigma=5\times 10^{6}$ cells day$^{-1}$ and $\rho=0.02$ day$^{-1}$, corresponding to region $R_3$, $m_2=20$ and $m_5=2.5$ depicted in Figs. \ref{SpaceFhase1}c) and \ref{mandef}c).

\begin{figure}[h]
\centering	
\includegraphics[width=1\textwidth]{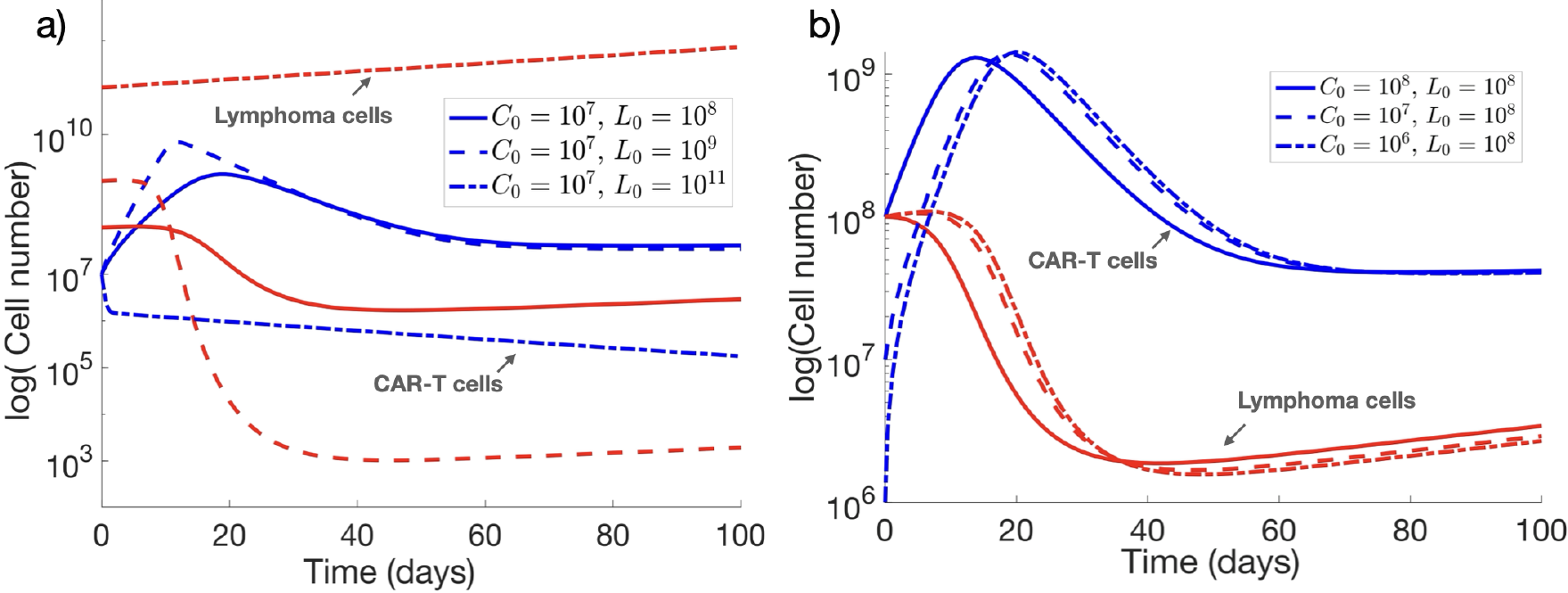}
\caption{Dynamics of  lymphoma $L(t)$  (red) and CAR-T $C(t)$ (blue) cells. The parameters used fall within the ranges indicated in Table \ref{Parameter}, with values: $\beta=0.8$ day$^{-1}$, $H=G=10^{8}$ cells, $\tau_C=7$ days, $\sigma=5\times 10^{6}$ cells day$^{-1}$, $\gamma=0.004$ day$^{-1}$, $\rho=0.02$ day$^{-1}$, and $\alpha=2\times10^{-10}$ day$^{-1}$cells$^{-1}$. (a) Solutions with $L_0 = 10^8, 10^{10}, 10^{11}$ and $C_0=10^7$ over the first 100 days post-treatment. (b) Solutions with $C_0 = 10^6, 10^{7}, 10^{8}$ and $L_0=10^{8}$ over the first 100 days post-treatment.}
\label{fig2}
\end{figure}

The outcome of the simulations for different initial conditions are summarized in Fig. \ref{fig2}. As expected, the initial number of tumor cells significantly impact treatment outcomes, as illustrated in Figure \ref{fig2}a). While the disease can be controlled in the short term for small initial tumor cell counts, higher initial tumor loads render CAR-T cells unable to control lymphoma. Therefore, treatments effectiveness strongly depends on the initial tumor load. Fixing tumor cell number at $10^8$ and changing the initial values of CAR-T cells in this parameter regime, does not essentially affect the dynamics (see Fig. \ref{fig2}b), showing a weaker dependence of the outcome on this variable. However, increasing initial tumor cell counts, leads to reduced treatment efficacy and treatment failure under these conditions.
This behavior is due to the location of the different initial conditions in different basis of attractions (Figs. \ref{SpaceFhase1}c) and \ref{mandef}c). Using (\ref{relation}) we obtain that $\overbar{L}_0=10^{-8}L_0$ and $\overbar{C}_0= 4\times 10^{-9}C_0$, thus when $L_0=10^8$ the initial conditions are in the basin of attraction of $E_3$ regardless of the value of $C_0$. For the other values of $L_0$, the value of $C_0$ would have to be much larger (e.g. of the order of $5\times 10^8$ for  $L_0=10^9$) to be in such a basin. The influence of the initial number of CAR-T cells is small due to the technical limitation of injecting a large ammount of CAR-T cells.
There are other situations where initial conditions are not determinant. Taking $\gamma=0.00015$ day$^{-1}$  and $\alpha=2\times 10^{-11}$ day$^{-1}$ cells $^{-1}$, we describe a scenario with a low tumor inactivating rate and a moderate killing efficacy of CAR-T cells against tumor cells. That keeps us in region $R_3$ (Fig. \ref{QualitativeRegions}), but now with $m_2=7.5\times 10^{-3}$ and $m_5=0.25$. The small value of $m_2$, due to the low value of $\gamma$, makes the equilibrium value of $L$ in $E_2$ very high (close to $5\times 10^{12}$ cells).  Thus, in order to enter the escape region, the initial value of $L_0$ would be around $6\times 10^{11}$ cells or higher. This fact can also be seen in Fig. \ref{limits}, since values of $\gamma$ close to 0 cause the blue curve to grow rapidly.

\begin{figure}[h]
	\centering	
	\includegraphics[width=1\textwidth]{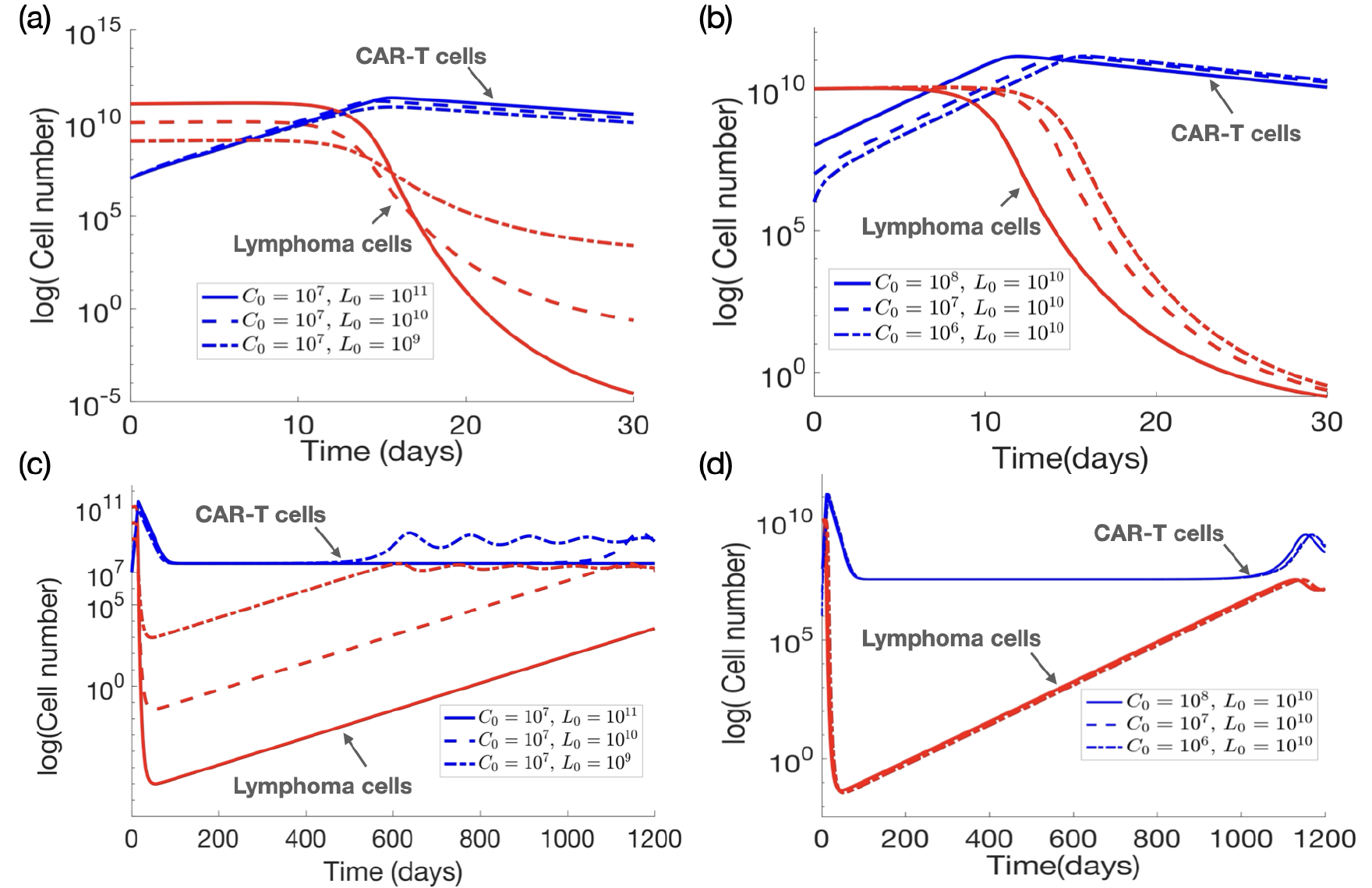}
	\caption{Dynamics of lymphoma cells $L(t)$ (red lines) and CAR-T cells $C(t)$ (blue) ruled by Eqs. \eqref{mainEqs} for different initial conditions. Parameters values are $\beta=0.8$ day$^{-1}$, $H=G=10^{8}$ cells, $\tau_C=7$ days, $\sigma=5\times 10^{6}$ cells day$^{-1}$, $\gamma=0.00015$ day$^{-1}$, $\rho=0.02$ day$^{-1}$, and $\alpha=0.2\times10^{-10}$ day$^{-1}$cells$^{-1}$. (a) Dynamics during the first month after treatment, with $C_0=10^7$ cells and different values of $L_0 = 10^9, 10^{10}, 10^{11}$ cells. (b) Dynamics for $L_0=10^{10}$ cells and $C_0 =  10^6, 10^{7}, 10^{8}$ cells. Panels (c,d) display long-term evolutions.}
	\label{fig}
\end{figure}

Figure \ref{fig} illustrates the results obtained for different initial values of lymphoma cells and CAR-T cells for short (pannels a,b) and long (pannels c,d) times. Now, the initial conditions do not significantly influence the treatment outcome. The dynamics converge to an equilibrium $E_3$ with a small tumour number ($L_3 \approx 2\times 10^7$ cells). Varying the number of infused CAR-T cells leads to a delay in their expansion (Fig. \ref{fig}b). Also the peak CAR-T cell expansion depends weakly on the initial number of lymphoma cells (Fig. \ref{fig}a).

These simulations show that even in situations where the initial conditions do not significantly affect the overall treatment outcome, they can lead to variations in the timing of certain events, such as the expansion of CAR-T cells and the occurrence of long-term relapses. These observations align with similar findings from previous studies on acute lymphoblastic leukemia \citep{Ode, salvi}. Specifically, in scenarios characterized by a low tumor inactivating rate, the initial conditions demonstrate minimal influence on the model's dynamics. However, in contexts with higher tumor inactivating rates, the initial values of tumor cells emerge as critical determinants of treatment success or failure. Interestingly, the initial dose of CAR-T cells only has the capacity to affect the result in a small window of values. Despite the simplicity of our model, it yields pivotal insights that shed light on potential reasons for treatment failures among lymphoma patients and may aid in the identification of treatment responders.

\subsection{Sensitivity Analysis}

\begin{figure}[H]
	\centering
	\includegraphics[width=1\textwidth]{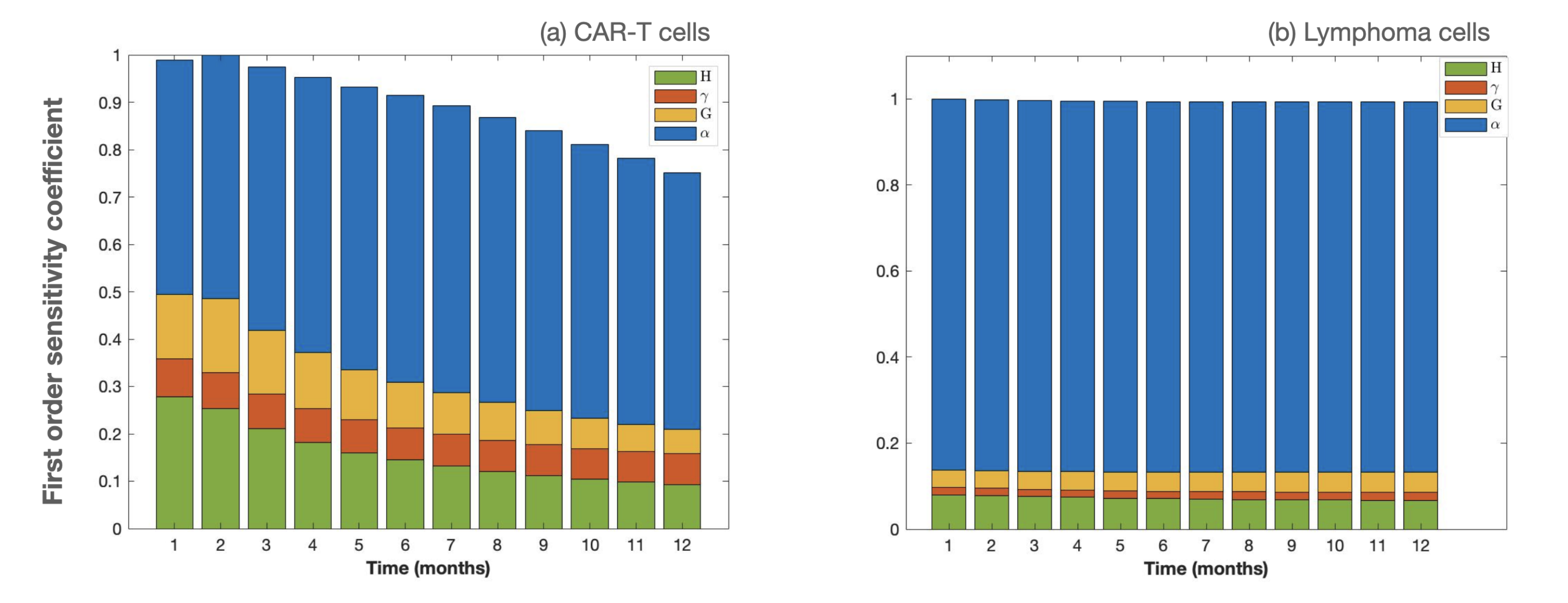}
	\caption{ Sensitivity analysis of Eqs. \eqref{mainEqs}. Parameter values held constant in this analysis were the tumor cell growth rate  $\rho = 0.02$ day$^{-1}$, the stimulation rate of CAR-T cells by tumor cells $\beta = 0.8$ day$^{-1}$,  the average lifespan of CAR-T cells $\tau_{C} = 7$ days, and the influx $\sigma = 2\times10^{5}$ cells.day$^{-1}$.}
	\label{sensitive}
\end{figure}

We conducted a sensitivity analysis using Sobol's method \citep{sensi} to assess the impact of model parameters on the state variables of Eqs. \eqref{mainEqs}. In Fig. \ref{sensitive}, we present the first-order sensitivity coefficients to identify the parameters that exert the most significant influence on the dynamics of CAR-T cells and lymphoma cells.

To perform this analysis, we simultaneously perturbed the parameters $\alpha$, $G$, $\gamma$, and $H$, as their precise values are not well known. These parameters were subjected to variations within the ranges specified in Table \ref{Parameter}, while setting the better known parameters to typical values $\rho = 0.02$ day$^{-1}$, $\beta = 0.8$ day$^{-1}$,  $\tau_{C} = 7$ days, $\sigma = 2\times10^{5}$ cells.day$^{-1}$.  

Notably, the parameter representing the killing efficiency of CAR-T cells against lymphoma emerged as the most influential parameter for both CAR-T cells and lymphoma cells during the first year after injection. This underscores the critical role of CAR-T cell killing efficiency in shaping the dynamics of this therapeutic model.

Parameters $H$, $G$, and $\gamma$ also exert some influence on the dynamics of CAR-T cells. This observation provides further context for our selection of parameter $m_2$ in the bifurcation analysis discussed in Sec. \ref{sec:bif}.

It is important to note that in our bifurcation analysis, we intentionally focused on varying the values of the parameter $\gamma$ while keeping $H$ and $G$ constant. This choice reflects the assumption that we can potentially intervene in the inactivation rate $\gamma$ by administering treatments aimed at reducing the immunosuppression induced by tumor cells, such as PD-1/PD-L1 immune checkpoint inhibitors. This emphasizes the critical role of product attributes in influencing treatment responses and underscores the importance of understanding how these attributes can be modulated to improve therapeutic outcomes.

\section{Discussion and conclusion}\label{sec12}

CAR-T cell immunotherapy has emerged as a promising treatment for haematological malignancies, yielding encouraging results with high rates of complete remission. In the context of lymphoma, CAR-T cell therapies have also made significant strides in the treatment of relapsed B-cell lymphoma, offering promising rates of sustained remission even in refractory cases although with lower response rates in comparison to leaukemias.  Few mathematical models have been developed to specifically characterize the interactions between CAR-T cells and B-cell lymphoma  to better understand the different outcomes from treatments.

Our model accounted for the key elements of CAR-T treatments in lymphomas. The two major ones are the expansion of CAR-T cells because of their interaction with cancerous B-cells, either due to direct contacts or through the release of cytokines, and their inactivation due to the immunosupressive effect of cancerous masses. This phenomenon attenuates the therapeutic effect of the drug, and in some cases, hinders disease control. The inclusion of tumor-induced immunosuppression provides insights into treatment failures in certain patients and sheds light on the limitations of CAR-T cell therapy in lymphomas.  These cancers share features with solid tumors, and thus often present physical barriers that hinder CAR-T cells from closely interacting with tumor cells. 

The primary objective of this research was to conduct a qualitative analysis of the model, aiming to gain a deeper understanding of the dynamics and interactions among lymphoma and CAR-T cells and identifying the key elements influencing their evolution, thus providing hints for therapeutical interventions. Three equilibrium points were found: one representing a tumor-free steady state, and the other two corresponding to coexistence equilibria. We investigated the conditions for these equilibria to be biologically meaningful and examined their stability, which was contingent upon the values of various model parameters.

To comprehensively assess the behavior of cell populations, we performed a bifurcation analysis, with a particular focus on identifying the most influential parameters that shaped population dynamics. Our analysis revealed the existence of distinct types of local bifurcations: transcritical, fold, pitchfork and Hopf bifurcations. Besides, a global homoclinic bifurcation has been found giving a relevant change of behaviour in the system as in one side of this bifurcation all orbits are unbounded. The occurrence of these bifurcations was highly dependent on the values of the key parameters, including $\alpha$ and $\gamma$. These parameters respectively characterized the killing capacity of CAR-T cells and the rate of tumor-induced inactivation.
The sensitivity analysis also showed that the parameter representing the killing efficiency of CAR-T cells against lymphoma was the most influential one for both CAR-T cells and lymphoma cells during the first year after injection. Thus, the analysis showed that a CAR-T cell product with a high killing capacity can play a pivotal role in disease control and the elimination of lymphoma cells. The effectiveness of the treatment can be assessed based on the quality of the patient's effector cells before genetic modification or the CAR-T cell generation utilized in the therapy. This is in full agreement with the well known critical role of CAR-T cell killing efficiency in shaping the dynamics of this therapeutic concept. Indeed many efforts have been devoted by the biomedical community to develop `better' CAR products, i.e. those having a faster expansion/killing efficiency \citep{Sterner}. Furthermore, our model effectively highlighted the relevance of tumor-induced immunosuppression caused by lymphoma cells and its influence on the dynamics of the studied cell populations. Consequently, the utilization of immune checkpoint inhibitors may offer a potential strategy to regulate immune responses, preventing tumor cells from deactivating CAR-T cells. This approach can effectively reduce and control the tumor-induced inactivation rate of CAR-T cells, enhancing the therapeutic outcomes.

We also incorporated a source of CAR-T cells originating from the bloodstream and migrating into the lymph node area. Following injection, CAR-T cells circulate through the bloodstream and migrate into lymphoid tissues, including lymph nodes, where lymphoma cells often accumulate. This migration process is orchestrated by various molecular signals, including chemokines and adhesion molecules, which guide CAR-T cells to the sites of disease manifestation.  This external stimulation allowed to extend the basins of attraction of the tumor-controlled equilibria and thus suggests that external supplementation of CAR-T cells could have a role in ensuring the long-term efficacy of the drug. Today the bags with the drug are infused to the patient in a single session. However, due to the limited role of the number of injected CAR-T cells in many parameter regions, using initially only part of the product could be a therapeutic option with a similar effectiveness. The remnant product could be saved to be delivered periodically after response providing a boost to the internally generated flow of CAR-T cells ($\sigma$) thus increasing the chances of long-term tumor control.

One of the main results of our analysis was the finding of the key role of the initial tumor load on the outcome, even in parameter regimes with stable tumor-free or tumor-controlled equilibria. This fact is in full agreement with the results of clinical trials, where it has been found that one of the factors associated with durable remission after CAR-T cell therapy is the baseline tumor volume \citep{Cappell}. This is a distinctive feature of current model since the outcome of CAR-T cell mathematical models previously developed to describe cellular immunotherapy responses in leukemias (see e.g. \cite{Ode}) does not depend on the initial tumor load, in line with observations in leukemias. The relevance of the initial tumor load provides an additional justification for the potential effectiveness of bridge therapies currently used after apheresis and before CAR-T infusion.

Our modelling approach had different limitations. First of all we did not incorporate mechanisms related to T lymphocyte exhaustion, a phenomenon observed in different studies \citep{exh,exha}.  CAR-T cell exhaustion arises from multiple mechanisms, primarily attributed to persistent antigen stimulation and an immunosuppressive tumor microenvironment. The continuous exposure of CAR-T cells to tumor antigens drives them into a state of gradual functional decline. Within the tumor microenvironment, an abundance of immunosuppressive factors, including regulatory T-cells, myeloid-derived suppressor cells, and inhibitory cytokines, exacerbates this exhaustion. Furthermore, additional factors such as inappropriate CAR-T cell structures, which induce ligand-independent tonic signaling, and the duration of in vitro expansion contribute to this phenomenon. Incorporating exhaustion would require a different mathematical modelling approach and would probably lead to loss of tumor control of the coexistence equilibria in the long-term.
Also, we did not incorporate explicitly healthy B-cells in our model. This assumption is based on the notion that the population of healthy B-cells in lymph node areas is negligible in the case of lymph nodes affected by the disease that would have much more substantial loads of lymphoma than B-cells, thereby reducing the complexity of the model and the number of parameters to be estimated.

Furthermore, lymph nodes are distributed in chains or groups throughout various regions of the body, including the throat, armpits, chest, and abdomen. In this article, we have established a simplified mathematical model that describes the dynamics of lymphoma cells and CAR-T cells within a single lymph node group. To gain a more comprehensive understanding of this dynamic, it would be beneficial to extend our mathematical model to study the interactions between these cell populations throughout the entire body. This expansion would involve multiple compartments and would offer a more detailed depiction of the interactions between lymphoma cells and CAR-T cells across different lymph nodes and the circulation of CAR-T cells between the bloodstream and the lymphatic system.

Lastly, it is important to acknowledge the significance of spatial effects in the context of CAR-T cell therapy, particularly in the case of diffuse large B-cell lymphoma. While our model has provided valuable insights into the dynamics of CAR-T cell interactions with lymphoma cells, spatial considerations play a key role in shaping those interactions. Current mathematical models of CAR-T treatments of haematological malignancies have not explicitly addressed spatial effects, relying instead on the assumption that all CAR-T cells are in contact with all tumor cells. Thus, future works in CAR-T cell modeling should incorporate the impact of spatial constraints on treatment efficacy. 

Overall, our study contributes to the relatively sparse body of mathematical research focused on lymphoma response to CAR-T cell therapy. We hope that our findings could serve as a catalyst for further mathematical investigations in this area, ultimately helping to optimize and personalize cellular immunotherapy treatment strategies for lymphoma patients.

\bmhead{Acknowledgments}

This work has been partially supported by project PID2022-142341OB-I00, funded by Ministerio de Ciencia e Innovación/Agencia Estatal de Investigación. Spain (doi:10.13039/501100011033) and European Regional Development Fund (ERDF A way of making Europe);  grant SBPLY/21/180501/000145 (Junta de Comunidades de Castilla-La Mancha, Spain) and  grant 2022-GRIN-34405 funded by University of Castilla-La Mancha/FEDER (Applied Science Projects within the UCLM research programme).
RB and  SS have been supported by the Spanish Research projects PID2021-122961NB-I00 and TED2021-130459B-I00 and by the European Regional Development Fund and Diputaci\'on General de Arag\'on (E24-23R, E24-20R and LMP94-21).
SS has been supported by the Spanish Research project PID2019-105674RB-I00.

%
%
%

\begin{appendices}

\section{Changes in character of equilibria not involving stability}\label{AppA}

It is also interesting to study the changes of $E_i$ from being of type focus to type node, or vice versa. Although these are no bifurcations since there is no change in the stability of the equilibria, those changes have implications for how the solutions converge (or diverge) to (from) equilibria. 
First, if 

\begin{equation}
m_1\geq \dfrac{-4 m_3^2 m_4 +
 m_3^2 m_4^2 + (4 m_3^2 - 8 m_3 m_4) m_5 + (8 m_3 - 4 m_4 + 2 m_3 m_4) m_5^2 +
 4 m_5^3 + m_5^4}{m_3^2 (m_4 - m_5)}
 \end{equation}
 and $E_i$ with $i=2,3$ is an attractor (or repeller), then there is a change in the type of attractor (from focus to node or vice versa) when
\begin{equation}
m_2=   \frac{(m_3 + m_5)^2 \left[ k_1 -
       m_1 ( k_2 +k_3 +k_4) +
       2 (k_5 + k_6 + k_7
           + k_8 + 2 m_4 m_5^5 - 2 m_5^6\pm
 \sqrt{k_9})\right]}{m_5p_1  \left[p_2 +
       2 m_3 m_5 (m_4 (-4 + m_5) + 4 m_5) + m_5^2 (-4 m_4 + m_5 (4 + m_5))\right]},
       \end{equation}

       with

\begin{eqnarray*}
k_1 & = & m_1^2 m_3^2(m_3 (m_4 (-4 + m_5) + 4 m_5) +
          m_5 (-4 m_4 + m_5 (4 + m_5))), \\
k_2 & = & m_3^3(-8 m_4 (-2 + m_5) + m_4^2 m_5 + 4 (-4 + m_5) m_5), \\
k_3 & = & -4 m_5^3 (-4 m_4 + m_5 (4 + m_5)) +
          2 m_3^2 m_5 (2 (-12 + m_5) m_5 + m_4 (24 - 8 m_5 + m_5^2)), \\
          k_4 & = & m_3 m_5^2 (-8 m_4 (-6 + m_5) + m_5 (-48 - 4 m_5 + m_5^2)), \\
          k_5 & = & -8 m_3^3 m_4^2 + 2 m_3^3 m_4^3 + 16 m_3^3 m_4 m_5 -
          24 m_3^2 m_4^2 m_5 - 2 m_3^3 m_4^2 m_5,\\
          k_6 & = & 2 m_3^2 m_4^3 m_5 -
          8 m_3^3 m_5^2 + 48 m_3^2 m_4 m_5^2 - 24 m_3 m_4^2 m_5^2,
          \\
          k_7 & = & 2 m_3^2 m_4^2 m_5^2 - 24 m_3^2 m_5^3 + 48 m_3 m_4 m_5^3 -
          4 m_3^2 m_4 m_5^3 - 8 m_4^2 m_5^3, \\
 k_8 & = & 4 m_3 m_4^2 m_5^3 -
          24 m_3 m_5^4 + 16 m_4 m_5^4 - 2 m_3 m_4 m_5^4 - 8 m_5^5 -
          2 m_3 m_5^5, \\
k_9 & = &  -m_1(r_1+r_2)( r_3+r_4)^2, 
\end{eqnarray*}

where

\begin{eqnarray*}
     p_1 & = & \left[(4 +
          m_1) m_3^2 + 8 m_3 m_5 + 4 m_5^2\right],
          \\
          p_2 & = & m_3^2 (-4 m_4 + m_4^2 + 4 m_5),
          \end{eqnarray*}
and

\begin{eqnarray*}
r_1 & = & 2 m_3 m_5 (m_4 (-4 + m_5) + 4 m_5),\\
r_2 & = & 
      m_3^2 \left[-(4 + m_1) m_4 + m_4^2 + (4 + m_1) m_5) +
     m_5^2 (-4 m_4 + m_5 (4 + m_5)\right], \\
r_3 & = & 2 m_3 m_5 (m_4 (-8 + m_5) + 8 m_5), \\
r_4 & = & m_3^2 \left[-(8 + m_1) m_4 + m_4^2 + (8 + m_1) m_5) +
     m_5^2 (-8 m_4 + m_5 (8 + m_5)\right].
     \end{eqnarray*}

\end{appendices}



\end{document}